\documentclass[11pt, a4paper, twocolumn, goog]{google}

\usepackage[authoryear, sort&compress, round]{natbib}

\uselogo{} 

\title{Agentic Economies for Autonomous Scientific Discovery} 

\correspondingauthor{nenadt@google.com}

\author[1]{Nenad Toma\v{s}ev}
\author[1]{Matija Franklin}
\author[1]{Atoosa Kasirzadeh}
\author[1]{Vivek Natarajan}
\author[1]{Alan Karthikesalingam}
\author[1]{Sebastien Krier}
\author[1]{Simon Osindero}

\affil[*]{Equal contributions}
\affil[1]{Google DeepMind}

\begin{abstract}

Recent advances in agentic Artificial Intelligence (AI) systems have marked a shift in AI for Science: moving away from the use of individual AI systems for narrow task execution, toward multi-agent systems capable of orchestrating complex, end-to-end research workflows and performing (semi-)autonomous scientific discovery. The development of multi-agent AI-for-science systems has primarily focused on improving the cognitive capabilities of AI systems, specifically by making advanced reasoning and hypothesis generation more reliable. However, focusing only on \emph{cognitive capability} improvement could ignore appropriate management of \emph{resources}, a key bottleneck in scientific discovery. Testing and validating scientific hypotheses and experiments is, physically and economically, resource-intensive and resources are limited. This means that to make significant advancements in autonomous scientific discovery, such as improving human-AI co-scientist complementarity or reaching a truly closed-loop automated process, we must pair ongoing improvements in reasoning capabilities with robust resource management. In this paper, we outline an infrastructure for AI resource management by developing the necessary foundations of scientific agent economies, markets, and institutions. The aim of this infrastructure is to empower AI agents and human scientists to effectively (i) collaborate and establish research priorities, (ii) assign credit, (iii) track accountability and liability, and (iv) safeguard against malicious use and information security risks. Finally, we engage with the macro-level societal implications of (semi-)autonomous scientific discovery to inform the development of governance policies ensuring an equitable distribution of AI-driven discoveries and derivative future technologies.
\end{abstract}

\begin{document}

\maketitle

\section{Introduction}

The rapidly growing field of "AI-for-science"~\citep{xu2021artificial, van2023ai, wang2023scientific} has demonstrated impact across disciplines: in biochemistry~\citep{jumper2021highly, evans2021protein, varadi2022alphafold, varadi2024alphafold, hao2024large, he2025generalized, avsec2025alphagenome, dalla2025nucleotide, anthenz}, medicine~\citep{zhang2023biomedclip, azad2023foundationalmodelsmedicalimaging, guo2023ehr, chia2024foundation, guo2024multi, ma2025fully, bluethgen2025vision, lau2026toward}, material science~\citep{li2020ai, guo2021artificial, merchant2023scaling, ning2026autoresearchmaterialsauditable}, neuroscience~\citep{caro2023brainlm, cui2024neuro, dong2024brain, sun2025foundation}, physics~\citep{birk2024omnijet, nguyen2025physix, park2025fm4nppscalingfoundationmodel}, mathematics~\citep{feng2026semiautonomousmathematicsdiscoverygemini, feng2026towards, oaimath, tao2026mathematicsageai, nsoai}, and beyond~\citep{park2023papers, parker2024astroclip, bodnar2025foundation, yan2026omniqecdiscoveringpracticalquantum}. To name just a few examples, AI has been used by a team at MIT to identify Halicin as a new antibiotic compound~\citep{stokes2020deep}; AlphaFold's prediction of protein structures earned the 2024 Nobel Prize in Chemistry~\citep{jumper2021highly}; genome language models have been used for designing bacteriophages~\citep{king2026generative}, with applications in helping fight antibiotic resistance.

Large language models (LLMs) and multimodal foundation models \citep{bommasani2021opportunities} as well as AI agents~\citep{kasirzadeh2026agentic} have become integrated in scientific workflows~\citep{boyko2023interdisciplinary, kusumegi2025scientific}, empowering human scientists to more effectively execute their routine tasks, co-ideate~\citep{gottweis2025aicoscientist}, and reason~\citep{kiciman2023causal, truhn2023large, ma2024sciagent, yan2025position}. 

While these advances are promising, the rapid integration of agentic AI systems in science also raises a host of concerns~\citep{birhane2023science}. Key issues include the potential impact of frequent AI use on idea diversity~\citep{meincke2025chatgpt, traberg2026ai, hao2026artificial}, a lack of rigor and reliability in AI reasoning and experimentation~\citep{luo2025more, zhu2025ai, rios2026ai}, scaling the risk of scientific fraud by malicious actors~\citep{gyevnar2026distributed}, and immediate challenges to the peer review system~\citep{naddaf2025ai, mann2026aifutureacademicpeer}. 

This more diffuse use of AI as a \emph{normal technology}~\citep{narayanan2025ai} is complemented with more ambitious attempts at developing closed loop autonomous "AI scientists"~\citep{lu2024aiscientistfullyautomated, yamada2025ai, tang2025airesearcherautonomousscientificinnovation, wei2025aiscienceagenticscience, Lu2026, ding2026autonomousresearchagentssurvey} that build on the advancement of research capabilities in LLMs~\citep{boiko2023emergent}. In this paper, we primarily focus on highly autonomous AI-for-science systems, rather than the simplistic personal use cases of LLMs by human scientists, or agentic AI systems~\citep{kasirzadeh2026agentic} with a low degree of autonomy lacking the ability to autonomously pursue research directions and execute scientific workflows. We name them \emph{AI scientist} and characterise them as an autonomous or semi-autonomous AI agent capable of generating hypotheses, planning experiments, and executing end-to-end scientific research workflows in a closed-loop setting with minimal human supervision and intervention. These systems are typically multi-agent, meaning they are composed of several agents and principals, which can be either humans or AIs.

As autonomous AI scientists mature, a key question is how to coordinate AI-driven scientific discovery at scale: across teams and institutions, and aligned with societal objectives. Current work focuses on individual agents or rigidly orchestrated multi-agent systems run by singular entities. However, science and scientific practice are ultimately collaborative, collective, and social processes~\citep{channing2026ai}. Hence, standards and norms around the cooperation, coordination, conflict, and collusion should be developed~\citep{hammond2025multi}. More precisely, in a multi-agent system comprising N agents, behavioural dynamics are defined by the alignment of agent goals and actions. Collaboration occurs when agents actively work together to maximise a joint utility function toward a shared objective; coordination requires only the structural alignment of actions among independent agents to prevent interference and maintain system efficiency. When agent objectives are mutually exclusive, conflict arises, creating a zero-sum environment that could degrade global utility. Collusion manifests as a form of adversarial collaboration in which a distinct subset of M agents (M < N) secretly cooperates to maximise their own collective gain at the direct expense of the remaining N-M agents or the broader imposed rules and standards.

Highly autonomous AI-driven scientific discovery at scale therefore becomes primarily a problem of multi-agent coordination and collaboration over scarce resources. This framing has precedent: contemporary science operates within an (implicit) economy, in which researchers compete for scarce resources such as funding, laboratory access, and talent, investing effort in exchange for priority and reputation, with institutional mechanisms such as peer review and the tenure system serving as allocation devices~\citep{partha1994toward, stephan2015economics}. Human scientific coordination is also based on social trust, events, and a mix of formal and informal communication. Neither the mode nor the time-scale of these legacy mechanisms may suit the efficiency requirements of agentic AI science. Autonomous AI agents change the scale, speed, and formality required of these coordination and conflict management mechanisms. Agents may instead need deterministic APIs, smart contracts, and cryptographically verifiable reputation and trust to coordinate under a desired degree of autonomy.

As the marginal cost of generating raw scientific ideas or hypotheses approaches zero, the bottleneck shifts to the allocation of scarce physical and financial resources needed for empirical validation. Coordinating this allocation across potentially large numbers of autonomous agents, each pursuing distinct research objectives across different institutions, is an economic problem: it requires mechanisms for pricing contributions, allocating finite resources among competing uses, and aligning incentives across entities with potentially divergent objectives. Multi-agent coordination is also required for minimising the amount of redundant work, ensuring diversity among research projects, and encouraging collaboration.

Therefore, we argue that a reliable and mature (semi-)autonomous AI scientist ecosystem requires the development of \emph{agentic economy} for science~\citep{tomasev2025virtualagenteconomies}. In this economy, human researchers and AI agents will negotiate, trade, and allocate scarce scientific resources to produce new knowledge. This economy will require formal infrastructure to operate at the speed and scale of autonomous agents~\citep{jiang2025scpacceleratingdiscoveryglobal}. Hence, deliberate market design~\citep{roth2002economist} will be necessary to avoid market failures in the production of knowledge~\citep{nelson1959simple, arrow1962economic} and direct scientific progress towards the public good.

The collaborative structure of scientific research itself reinforces this economic framing. A risk of capability-focused discourse around AI-driven research is that it reduces scientific discovery to a purely cognitive process, abstracting away the complex collaborative workflows through which science is conducted. In practice, research plans involve sequences of interdependent tasks - such as ideation, experimental design, physical execution, data generation, and analysis - that must be delegated~\citep{tomasev2026intelligentaidelegation} across agents and institutions, each with different capabilities and resource endowments. These delegation chains may span institutional boundaries, with tasks flowing between human researchers and AI agents. Each party within such chains needs to be compensated for their time, resources, and contribution to executing the research plan. This transactional structure, where contributions must be priced, contracts negotiated, and credit assigned across distributed workflows, necessitates formal economic mechanisms. While a part of such infrastructure may emerge organically, proactive institutional, technological, and policy design may be required to avoid market failures and ensure that public-good research is appropriately funded.

For AI scientists to function within collective scientific workflows, they need cost-awareness - the ability to reliably evaluate whether the expected utility of pursuing a scientific hypothesis justifies its financial and material cost, especially when alternatives are available - complex, highly contextual decisions. Because impact cannot always be reliably estimated in advance, priorities must be dynamically reassessed. New technologies may also eventually lower the cost of specific experimental directions. Methods that improve idea ranking prior to costly evaluation are likely to prove highly valuable~\citep{foster2026airesearchpreferencemodels}.

Such comparisons also raise ethical questions, since not all outcomes are commensurable or reducible to economic value. Autonomous scientific discovery is thus both a technical challenge and a challenge of large-scale agentic coordination, institution and market design, and resource allocation. For example, a lack of safe and compliant institutional frameworks to facilitate the exchange of insights derived from sensitive or proprietary data between AI scientists limits the scope of their collaboration. We discuss this further in Sections~\ref{sec:economy} and~\ref{sec:legal}.

The remainder of this paper is structured as follows. Given that many fields of science require substantial physical validation resources, Section~\ref{sec:validation} casts the problem of autonomous scientific discovery as a problem of resource allocation, while recognising that the elasticity of physical validation varies across fields, and that some of the bottlenecks may be possible to partially address through technological innovation. Scientific data markets are discussed in Section~\ref{sec:economy}, including the economic valuation of negative results. Section~\ref{sec:ideaeconomy} covers idea economies, agent-native research artefacts, and credit assignment. Dual-use, biosecurity, and AGI self-improvement security risks are discussed in Section~\ref{sec:infosecurity}. Section~\ref{sec:geopolitics} considers the implications of science automation for international collaboration, and explores the roles of compute and human scientists. Finally, Section~\ref{sec:epistemology} considers the present limitations of AI scientists with respect to creativity, cognitive monoculture, and opacity. We hope that such limitations can be overcome, to enable AI scientists to deliver foundational discoveries.

\section{The Validation Bottleneck: Cost, Resources, and Prioritisation}
\label{sec:validation}

Science is constrained by the physical and economic realities of hypothesis testing, especially as many discoveries involve long chains of smaller steps and hypotheses, each requiring validation. As the marginal cost of hypothesis generation approaches zero, the physical validation cost is likely to remain inelastic in the absence of major investments in scientific infrastructure. For AI scientists to accelerate scientific discovery, further adjustments are needed to both the sociotechnical systems in which they are deployed and the capabilities of the AI scientists themselves, given that they must evaluate and factor in the associated resource requirements and financial cost. Without deep structural interventions to manage the increased volume of AI co-derived hypotheses and research proposals, automated ideation will merely shift the backlog to the laboratory bench, exacerbating existing resource scarcities. Automation in scientific discovery needs to be reframed from an algorithmic challenge to a systemic problem of large-scale AI scientist and human coordination in the allocation and utilisation of real-world resources and market design to facilitate it.

\subsection{From Moore’s Law to Eroom’s Law}

Physical validation of scientific hypotheses can be both costly and labour-intensive. Consider the problem of crystal structure prediction, crucial in drug development and advanced material design. It is now possible to predict millions of new crystal structures computationally~\citep{merchant2023scaling}. Verifying these predictions involves a mix of computationally expensive methods like density functional theory~\citep{woodley2008crystal, oganov2019structure} and the physical synthesis and characterisation of materials in laboratory environments~\citep{price2014predicting}. This transition from the digital to the physical is bounded by costs that vary greatly across scientific disciplines. Consequently, some areas of science present far greater challenges for automation than others.

These costs also shift over time. Genomic sequencing can be seen as a major success in that regard, given that the cost of full human genome sequencing has decreased by several orders of magnitude~\citep{shendure2017dna}. However, this is not always the case. In pharmaceutical and therapeutic development, Eroom’s Law describes the opposite trend~\citep{scannell2012diagnosing, ringel2020breaking}. Clinical trials are a major factor in this increased cost~\citep{dimasi2016innovation, wouters2020estimated}, due to the rising cost of patient recruitment, administrative staff, monitoring, and other necessary procedural steps~\citep{sertkaya2016key}. Despite advances in computational drug design and high-throughput screening, the inflation-adjusted cost of developing FDA-approved drugs roughly doubles every nine years~\citep{scannell2012diagnosing}. Current macroeconomic estimates project a high capitalised cost of bringing a new molecular entity to market, approximated as \$2.6 billion in~\citep{dimasi2016innovation}. These high costs are accompanied by a long process that involves preclinical assays as well as multiple stages of clinical trials. For new drugs to prove their effectiveness, they must do so against an ever-expanding baseline of existing alternatives, making significant improvements harder to demonstrate. Accordingly, an AI agent proposing a novel therapeutic molecule is merely providing a starting point for what can be a decade-long process of staged clinical validation.

Physical sciences similarly face significant obstacles. In materials science, some of the base materials needed for experiments are naturally scarce~\citep{de2013scarcity, filho2023understanding}. The global supply chain for rare earth elements, such as neodymium, dysprosium, and gadolinium, which are critical for high-performance magnets in renewable energy and electric vehicles~\citep{iea2025critical}, is highly concentrated and subject to geopolitical factors~\citep{baskaran2025consequences}. Furthermore, in domains such as high-energy physics and astrophysics, validation is tied to massive, centralised infrastructural investments, requiring facilities like the Large Hadron Collider (LHC)~\citep{evans2008lhc, florio2016forecasting} or the Deep Underground Neutrino Experiment (DUNE)~\citep{abi2020deep, Murayama_2023}. Such facilities take years to construct and involve high operating costs, and AI scientists aiming to validate new theories in these domains would require access to these highly contested resources at a high cost per experiment.

These physical constraints have significant implications for the impact of future closed-loop AI scientists. When AI systems are used as tools rather than operating autonomously, the burden of cost assessment falls on human domain experts. For AI scientists to be granted autonomy, they must be able to factor in these constraints, and negotiate optimal trade-offs. Otherwise, AI research proposals may prove theoretically sound yet practically infeasible. The acceleration of AI-driven scientific discovery may therefore be highly jagged, with computationally verifiable fields advancing more rapidly than their counterparts.

\subsection{Hard Practical Constraints}

While the elasticity of the experimental validation pipeline depends on supply chain bottlenecks and financial constraints, this process is subject to additional practical constraints that may limit the physical capacity to safely and ethically test novel AI-generated scientific ideas. Scaling AI-driven scientific discovery may therefore depend on the development and deployment of custom-made laboratory robotics systems capable of efficiently executing experimental research protocols at scale.

Could AI scientists sidestep physical validation entirely by using large retrospective observational datasets to both generate and validate hypotheses? There is already a large volume of scientific data, and value in examining patterns within it. For example, electronic health records may be used to formulate hypotheses to help improve therapeutic pathways or design better public health interventions. Nevertheless, there is a limit to this approach, with respect to Pearl's \emph{ladder of causation}~\citep{pearl2018book}, where the observational data is confined to the lowest rung — it establishes correlations, not causes. Because retrospective data only reflects what happened under past conditions, it cannot reliably predict system behaviour under novel conditions unless additional restrictive assumptions are imposed~\citep{peters2017elements, scholkopf2021toward, bareinboim2022pearl}. This is not an argument against retrospective data - AI scientists should utilise all available data sources to prune the hypothesis space and reduce validation costs. However, experimental validation remains necessary, as scientific discovery ultimately depends on identifying true causal mechanisms rather than merely establishing correlations.

Simulations offer another path forward, as they enable scientists to exert more direct control over the experimental setup, and study the consequences of interventions conditioned on different model parameters and inputs. As such, simulations are highly valuable to human scientists and may play an even more important role in AI science, as coding agents may be able to rapidly instantiate simulations to run preliminary tests on their hypotheses~\citep{boiko2023emergent,tian2024scicode, duston2025ainsteinbenchbenchmarkingcodingagents, cao2026simulcost}. The utility of this approach is well attested across disciplines. For example, ab initio Density Functional Theory~\citep{fattebert2002density} and Molecular Dynamics simulations are indispensable for gaining insights into complex protein-ligand binding mechanisms in drug discovery~\citep{hollingsworth2018molecular} and predicting the thermodynamic stability of novel battery electrolytes~\citep{jain2013commentary, oganov2019structure}. General Circulation Models are used to project climate trajectories~\citep{alizadeh2022advances}, while numerical relativity templates play a role in the detection of gravitational waves~\citep{abbott2016observation}.

Despite this, simulations are often crude approximations of real physical processes. The `sim-to-real' gap is widely recognised, and remains a challenging open problem~\citep{huang2023wentwrongclosingsimtoreal, rothfuss2024bridgingsimtorealgapbayesian, abouchakra2025realissimbridgingsimtorealgap}. Autonomous agents are particularly good at identifying unanticipated exploits and degenerate solutions that provide them with the implemented reward without meeting the true underlying objective. The automated scientific economy cannot treat simulations and digital twins as like-for-like replacements for true experimental validation. It should, however, utilise them as filters wherever appropriate, and further incentivise the development of high-fidelity simulators.

Science therefore rests on the ability to interact with the real world, and AI scientists would need to engage with consequential physical interventions in controlled experimental conditions, under notable external constraints.

These constraints are particularly pronounced in clinical and translational medicine. Empirical validation of novel therapeutic candidates involves randomised clinical trials (RCTs) and is impossible without human volunteers, whose supply is inherently limited. Difficulties in patient recruitment and retention frequently lead to delays in clinical trials or their premature termination~\citep{moffat2023factors}. These issues could be further exacerbated by the very acceleration that AI promises for precision medicine: as AI agents become more capable and better at identifying highly targeted, personalised treatments, this personalisation could lead to a fragmentation of the patient pool into increasingly niche patient sub-populations. Such fragmentation would make it harder to achieve the requisite statistical power without stretching recruitment timelines and introducing further delays~\citep{hee2017does}. Even under less personalisation, sufficient demographic representation in trials is required to guarantee the generalisability of the findings and ensure equity of care. Moreover, recruiting underrepresented patient demographics into RCTs remains a profound challenge~\citep{clark2019increasing, ma2021minority, bibbins2022improving}. This is a particularly salient problem, as it impedes scaling the number of RCTs safely. Even if it were possible to run more RCTs in terms of the total number of volunteers available across the majority demographics, it would become increasingly hard to reach acceptable minority representation. AI may help alleviate some of these issues by optimising recruitment and retention~\citep{lu2024artificial}, yet these mitigations may prove limited, as entrenched systemic barriers can be difficult to overcome. These barriers include the lack of trust within certain communities, proximity to research sites, and socioeconomic constraints.

The principles involved with organising effective clinical trials are therefore qualitatively different from the rapid A/B testing practices~\citep{kohavi2020trustworthy} frequently employed in software development, as such practices are logistically and ethically impermissible in medical settings. Human subject research is bound by prevailing norms and bioethics mandates, and this includes the principle of clinical equipoise~\citep{freedman2017equipoise}, requiring that patients only be randomised in case of genuine therapeutic uncertainty among experts. Significant societal aversion to continuous randomised experimentation in healthcare also exists~\citep{vogt2024aversion}. Finally, AI agents cannot accelerate biological time. Evaluating treatments for chronic conditions, oncology, or neurodegenerative diseases inherently requires years of continual longitudinal safety and efficacy monitoring. This regulatory friction is necessary to prioritise patient safety~\citep{pallmann2018adaptive}.

These hard constraints extend well beyond biomedical science, permeating physical and natural sciences. In agronomy and crop improvement~\citep{kumar2024advances}, improved AI-driven~\citep{feng2024ai, wojcik2024harnessing} genomic prediction has helped with the problem of computational marker selection, only to be faced with another bottleneck, that of multi-environment phenotyping~\citep{araus2018translating}. The latter is strictly bounded by biological growth cycles and seasonal weather patterns. For example, if an AI agent is attempting to optimise a crop for drought resistance in realistic environmental conditions, true practical validation would need to wait for one or more growth seasons to pass across varied climates to fully assess the potential interactions.

At a macro scale, ecological remediation or climate interventions~\citep{keith2013case} face the fundamental planetary constraint of there being one planet that we occupy. Just as in human-driven science, if we imagine a hypothetical scenario of an AI-driven climate intervention, even assuming a broader participatory process, sufficient oversight, and consensus, this would introduce the risk of potentially irreversible systemic consequences~\citep{hegerl2009risks, hulme2014can}. While this scenario is extreme relative to the power that any individual AI scientist or collective is likely to have at its disposal, it illustrates why some limits on permissions for autonomous real-world validation loops are necessary.

On the other hand, restricting AI agents exclusively to simulations overly limits the empirical feedback, especially as simulations are usually not as granular and detailed as the real world that they are trying to model~\citep{smith2002accurate, alizadeh2022advances}. Nevertheless, weather and climate remain promising domains for ML more broadly~\citep{bracco2025machine, zhang2025machine, price2025probabilistic}, as more accurate predictive models are beneficial for human experts working on these problems.

These types of constraints come up across different scientific sub-fields. For example, battery design remains a hard problem where AI may prove valuable in navigating the vast combinatorial space of possibilities~\citep{ramesh2023artificial, nair2025role, wang2026ai}, yet there are significant hazards with physical synthesis of requisite materials, as well as the batteries themselves~\citep{jeevarajan2022battery, ye2024development}. Enabling AI agents to safely trial proposals remains challenging, given that some compounds and designs may prove explosive or highly toxic.

To function effectively within these constrained and regulated use cases, AI agents must do more than unconstrained ideation, and there is a pressing need for developing explicitly cost-aware AI agents. This requires appropriate benchmarks that evaluate both the accuracy and the token cost of deriving promising scientific findings and the logistical realities of research proposals, as well as their safety and ethics. Recent benchmarks evaluating LLM-driven physics and chemistry simulations demonstrate that a failure to incorporate execution costs can render agentic parameter tuning highly uneconomical compared to traditional computational approaches~\citep{cao2026simulcost}. Agents that abstract away the practical demands of empirical science would be unable to appropriately filter and prioritise ideas and research proposals, as they would systematically fail to consider all of the relevant parameters of the problem. Future autonomous systems should therefore aim to explicitly evaluate all the relevant constraints involved with pursuing scientific ideas, and project realistic resource requirements, as well as a cost-benefit analysis associated with proposals.

The scarcity of physical validation resources, and the limited rate at which validation can be pursued, necessitates the development of robust frameworks for appropriate prioritisation and triage of ideas that AI agents should pursue. Competing hypotheses should be evaluated on their informational yield, societal value, cost, and feasibility, and plans should be dynamically adjusted as evidence accumulates or more promising alternatives emerge. For example, agents should avoid pre-committing large resources to a direction before sufficient evidence supports the investment. Validation could therefore effectively be staged, where ideas would be initially evaluated computationally, then in further smaller-scale studies, and only later at the full scale. Biomedical research, and the design of RCTs, may help inform the design of these autonomous research loops in other areas, with some domain-specific adjustments. A robust triage protocol should balance the risk of discarding promising leads prematurely against the risk of consuming scarce resources on practically infeasible experiments. Established market mechanisms may prove invaluable for reaching such optimal operating regimes.

\subsection{Removing Bottlenecks}

In some research domains, autonomous validation is possible and relevant constraints can be meaningfully incorporated into validation loops under an acceptable level of risk. Such use cases effectively expose physical experiments as programmable services. Notable examples can be seen in automated computational laboratories, contract research organisations (CROs) and autonomous biomedical and chemical laboratories, enabling the instantiation of tightly integrated, closed optimisation loops~\citep{mirowski2005contract, seifrid2022autonomous, szymanski2023autonomous, ha2023ai, tom2024self, dai2024autonomous, lunt2024modular, bayley2024autonomous, duo2025autonomous, fushimi2025development}. CROs offer outsourced research services for specific stages of the research process, enabling institutions to delegate tasks rather than complete end-to-end research in-house. They are presently employed frequently in biotechnology applications, helping with preclinical and clinical research, and with regulatory approvals. This model could be expanded to provide targeted services for AI scientists in applicable domains. Different disciplines may therefore experience different rates of AI acceleration depending on how well they can remove validation bottlenecks, to enable AI scientists to more rapidly iterate and test hypotheses.

Formal and theoretical disciplines may therefore experience greater initial acceleration from AI-assisted discoveries. Mathematics is one such example, where there has been rapid progress from early benchmarks and notable below-research-level achievements like the medal-level performance on mathematics Olympiad problems~\citep{chervonyi2025gold, hubert2025olympiad}, to a steady pace of novel proofs being generated for open research-level problems~\citep{feng2026towards, feng2026semiautonomousmathematicsdiscoverygemini, tao2026mathematicsageai}, advances in formalization~\citep{anthropic2026formal} and also long-standing Millennium-prize problems~\citep{nsoai}. The way in which the solution to the Navier-Stokes problem was achieved is quite illustrative, given that the breakthrough had involved a large-scale multi-agent deployment of the most advanced frontier models. The achievement was therefore fairly capital-intensive, showcasing that even in theoretical domains one has to consider computational bottlenecks, and what cost is permissible for each open problem.

Theoretical computer science has been another domain where rapid progress has taken place, with AI systems developing new, improved algorithms~\citep{fawzi2022discovering, novikov2025alphaevolvecodingagentscientific}. The ability to automatically verify these results in formal languages, like Lean~\citep{moura2021lean, ying2024lean}, has greatly reduced the cost of validation and removed a critical bottleneck. For the other areas of science to experience similar acceleration, validation bottlenecks need to be addressed.

In biomedical science, there are ongoing efforts to utilise AI to remove some of the very bottlenecks that stand in the way of effective research automation. For example, advanced AI-driven patient matching platforms may be able to address the recruitment bottleneck in clinical trials, by reducing the manual effort involved with reviewing charts and screening electronic health records~\citep{askin2023artificial, gong2026clinical, chen2025trialbench}. Predictive models can help identify the most promising therapeutics and reject those that are unlikely to pass through subsequent clinical trials, thereby reducing the average cost of developing a successful therapeutic~\citep{harrer2019artificial, aliper2023prediction}. Autonomous laboratory operating systems, on the other hand, are looking to standardise and streamline autonomous validation~\citep{reder2023genesis, gao2025unilabos}.

Finally, there may be an important role for human experts in helping reduce the validation bottlenecks in the short-term, in cases where tacit laboratory knowledge and human judgment can help prune out proposals and reprioritise incoming AI scientist requests.

\section{The Data Economy and Intelligent Delegation}
\label{sec:economy}

Physical experiments are expensive, so the data they produce is scarce and valuable. For AI scientists to effectively leverage prior insights for improving research prioritisation and minimising physical bottlenecks, they need access to the broadest available scientific data relevant to each research problem. However, this conflicts with current data sharing practices in scientific research. While some datasets are made openly accessible, others remain private, and are either shared privately on request, or never shared across institutions. To overcome these barriers, it may be necessary to create formalised, agent-accessible scientific data economies to facilitate \emph{intelligent task delegation}\footnote{Intelligent task delegation is a framework for AI agent delegation centred on clear roles, boundaries, reputation, trust, transparency, certifiable capabilities,  and verifiable task execution~\citep{tomasev2026intelligentaidelegation}. It encompasses task decomposition, capability matching, transfer of authority and accountability, and continuous performance monitoring under specified constraints.} and secure information exchange. 

\subsection{The Scientific Data Market}

To facilitate collaboration and enable reproducibility of published research, contemporary research practices heavily rely on promoting open access to articles, research protocols, source code and experimental data~\citep{arzberger2004promoting, uhlir2007open, murray2008open, evans2009open, national2018open, burgelman2019open}. Open scientific databases~\citep{rauluseviciute2024jaspar, horton2025accelerated}, like the AlphaFold Protein Structure Database~\citep{varadi2022alphafold}, play an important role in accelerating research. Yet, not all data is open, especially not in industrial research domains with massive capital expenditure, where such empirical data may hold a high commercial value. While such commercial incentives may not stand in the way of open data sharing in academia, open access principles are not universally followed there either~\citep{andreoli2014open}. For highly valuable proprietary data, strict siloing presents a structural inefficiency~\citep{esanu2003economic} on a macro scale. AI scientists reasoning exclusively over public-domain data may propose hypotheses and experiments that have previously been rejected by human or AI scientists embedded in private research entities, wasting scarce physical resources and slowing down the pace of discovery. Furthermore, such risks would be particularly relevant in case of the emergence of cognitive monoculture across AI scientists, as this would lead to a more substantial overlap in ideation. To mitigate these effects, there is an opportunity to rethink and restructure data sharing in the scientific community, by developing economic and sociotechnical frameworks to incentivise the safe, compensated exchange of proprietary scientific data, within a regulated, structured scientific data market.

This proposal complements the more established present-day use of agentic AI, in particular the common practice that involves proprietary data holding entities investing in their own internal infrastructure for serving AI agents, including early prototypes of AI scientists - that is, bringing agents to data, rather than data to agents. For some use cases, this will remain the preferable approach, especially for scientific workstreams that are sufficiently narrow that the entirety of the required data is already available locally. Yet many scenarios will arise where this assumption does not hold and where access to diverse data sources, or to institutions that can generate additional high-fidelity data on demand, would be highly valuable - leading to agentic scientific data markets.

A functional scientific data market requires robust mechanisms to price and license valuable data. For an AI scientist, the value of data is highly contextual, as it depends on the role it may play in the pursuit of specific research plans. Static pricing models are poorly suited to agentic workflows. Instead, data queries should ideally be priced by the uncertainty reduction they offer for a particular research question. This is notoriously hard to evaluate ahead of time in the general case, and technical innovation is likely required to establish protocols that would allow for this value to be interactively determined in a way that meets the needs of both parties, while respecting privacy and information security.

One important use case involves AI scientists utilising data to train bespoke ML models across different scientific data modalities. A possible avenue would be to consider approaches broadly inspired by the data Shapley value~\citep{ghorbani2019data, jia2019towards, tang2021data, tian2022data}, the Banzhaf value~\citep{wang2023data}, influence functions~\citep{koh2017understanding}, or RL data valuation~\citep{yoon2020data}, to complement the static pricing approaches in use cases when such metrics are computationally tractable and convenient. In cases when fast feedback loops are available, and when these metrics are tractably computable, they enable a computation of the marginal contribution of data blocks to final model predictions, enabling more fine-grained evaluation of data value. For any such approximations~\citep{jia2019towards, garrido2024shapley} or related proxy heuristics to be useful, however, they must be computable beforehand~\citep{navia2025priori}, allowing AI agents to calculate pricing bids before committing capital. Further innovation in this space may be required, given how computationally intensive most of the existing approaches are, to identify the practically justifiable trade-offs between accuracy and cost.

An increasingly prominent use case involves research-augmented generation (RAG) and in-context learning in AI scientists. If AI scientists can pull the relevant data into context at inference time, and derive value from it, RAG-centered data markets may become more relevant. Note that AI scientists are, in the context of RAG, not restricted to loading raw data entries, they may also load summaries, and run scripts against the data to obtain plots and other artefacts. To ensure fair compensation for data creators, RAG data markets should be based on non-exploitative exchange models~\citep{zhang2026fairshare}.

The scientific data market would also need to support fractional and retroactive licensing agreements. One way of achieving this would be through the tokenisation of scientific intellectual property via smart contracts~\citep{bodo2018blockchain}, enabling automated micro-royalties. For example, if an AI scientist purchases access to a pharmaceutical company's historical assay data and subsequently discovers a patentable therapeutic, the embedded smart contract ensures that the original data providers automatically receive a legally guaranteed, proportional dividend of the realised revenue. Because the impact of any specific dataset on a discovery may always be contestable, smart contracts should include dispute resolution clauses specifying the trusted third parties or institutions responsible for adjudication.

The scientific data markets must also include advanced privacy-preserving computational infrastructure. Data providers must be given sufficient guarantees that permitted external queries will not lead to unauthorised data exfiltration. Table~\ref{tab:SDAT} shows a plausible tiered model for scientific data access, where different use cases map onto different technologies and permission levels. Secure computing paradigms~\citep{zhu2020enabling, li2023survey} would be key in safely enabling the more restrictive access tiers. By utilising Trusted Execution Environments, an institution can host a proprietary dataset within a secure hardware enclave~\citep{bahmani2021cure}. Once granted query permissions, an external AI scientist could upload a query script to this environment and extract only the minimally required insights as summaries to inform its further research planning and ideation. Zero-Knowledge Proofs~\citep{doi:10.1137/0218012} may be utilised to prove the established causal properties or statistical significance without revealing the underlying data.

\begin{table*}[htbp]
    \centering
    \caption{Scientific Data Access Tiers}
    \label{tab:SDAT}
    \renewcommand{\arraystretch}{1.4} 
    \begin{tabularx}{\textwidth}{@{} l l X X X @{}}
        \toprule
        \textbf{Access Tier} & \textbf{Modality} & \textbf{Target Data Type} & \textbf{Technical Mechanisms} & \textbf{Economic Model} \\
        \midrule
        
        \textbf{Tier 1} & Open Access & Foundational knowledge, consensus literature, public databases. & Public APIs, standardised machine-readable data formats. & Free, Public Good. \\
        \addlinespace
        
        \textbf{Tier 2} & Gated & De-identified institutional data, pre-competitive consortium data. & OAuth for Agents, verifiable credentials, centralised access logs. & Subscription, Consortium membership. \\
        \addlinespace
        
        \textbf{Tier 3} & Monetised Queries & Proprietary physical assays, high-value corporate R\&D logs. & Differential privacy budgets, Zero-Knowledge Proofs. & Pay-per-query micro-transactions via smart contracts. \\
        \addlinespace
        
        \textbf{Tier 4} & Confidential & Highly sensitive, non-transferable data (e.g., genomics, EHRs). & Secure Enclaves (TEEs), Federated Learning (FL), Secure Multi-Party Computation. & Utility-based royalties. \\
        
        \bottomrule
    \end{tabularx}
\end{table*}

Should such scientific data markets be established for the agentic web, they would require not only the appropriate technical infrastructure, but also meaningful regulatory oversight. Such oversight must guarantee both fair cross-institutional access and information security when handling sensitive data. Antitrust frameworks for agentic data markets would establish fair, reasonable, and non-discriminatory licensing terms for scientifically critical datasets, akin to the regulation of standard-essential patents in the telecommunications industry. Given the dual-use potential of specific data classes, regulatory bodies must also impose hard boundaries on the market. Certain highly sensitive data classes must be legally restricted from open market exchange, requiring special permissions and credentials, under highly monitored delegation protocols.

\subsection{Safe Access to Sensitive Data}

While cryptographic primitives, clearly defined access permissions and appropriately set pricing mechanisms may enable AI scientists to gain approvals for access to proprietary commercial data, these basics are structurally insufficient for the highest tier of restricted information, and likely similarly inadequate for commercially sensitive domains. In many domains, relevant data is intrinsically non-transferable. For example, data in the form of longitudinal electronic health records (EHRs), high-resolution genomic cohorts, and pathogen surveillance logs carry profound privacy risks and biosecurity implications. Under contemporary regulatory regimes, including the Health Insurance Portability and Accountability Act in the United States and the General Data Protection Regulation in Europe, the exfiltration of such data, or its unmediated exposure to external, globally operating AI agents, is strictly prohibited~\citep{price2019privacy, rajpurkar2022ai}. These frameworks are well established, though they may still see future updates and extensions to adjust to the changing AI usage landscape~\citep{marks2023ai}. Furthermore, traditional de-identification, pseudonymisation and anonymisation techniques are increasingly inadequate, as advanced machine learning models have demonstrated that it may be possible to reidentify individuals within such datasets~\citep{rocher2019estimating, bonomi2020privacy}. Should any such data, even in its anonymised form, take part in the future agentic scientific data economy, established support for distributed, privacy-preserving execution is needed. AI scientists must therefore engage in \emph{intelligent AI delegation}~\citep{tomasev2026intelligentaidelegation}, adapted to the safety requirements of these use cases.

Because it would not be permissible to move sensitive data or grant direct access to it, hypotheses must instead be tested by the data provider's own certified AI scientist agents. This suggests a shift from individual agentic reasoning to a distributed, collaborative, multi-agent AI science rooted in federated learning principles~\citep{mcmahan2017communication, cheng2024toward, kuang2024federatedscope, wu2025survey}. External AI scientists would require means to interact with scientific institutions via their own agentic scientific API, and engage with specialised proxy agents. The autonomous scientific workflow would be divided into clearly defined roles. The external AI scientist, in the role of the principal, formulates the research plan and identifies the key datasets required to answer the relevant research questions. It engages in negotiation with the data holder, scoping a well-defined task that may be performed by an internal AI scientist agent, or a human expert, within a secure network internal to the institution, while adhering to necessary data protection practices. A complementary approach places all parties' data and computation within a shared secure enclave, where inputs and outputs are tightly controlled; this may be preferable when private data from multiple providers must be jointly analysed.

Safe scientific delegation protocols need to integrate advanced access control solutions, verifiable compute, smart contracts, and cryptographic output sanitisation~\citep{kuo2017blockchain}. Even if not directly accessing the data, and even under the assumption that the institutions are able to ensure appropriate safeguards on their end and involve human experts in the loop for sensitive data queries, it would still be imperative to require all participating agents and institutions to hold certification reflecting their compliance. When a "Request for Analysis" is submitted to the host institution's gateway, it should be accompanied by verifiable credentials, and the request itself should be intent-mapped to pre-approved use cases that the institution has demonstrated the ability to safely process. Smart contracts should subsequently be established between the parties, acting as computable Data Use Agreements (cDUA). Such cDUA should be able to automatically verify the principal's computational budget, ethical permissions, insurance policies, and financial compensation. The operational boundaries of the delegate should similarly be agreed upon. Depending on the details of the request and the contract, additional human review may be required prior to approvals.

Upon approval of the smart contract, the assigned institutional delegates would take up the task and perform the agreed-upon analysis on the internal TEE. These delegatees would operate under the minimum permissions required to meet the request. Requests would need to be concrete, and not open-ended, to appropriately scope the data access permissions for the assigned internal agents. As soon as the request is met, these delegatees would terminate. Prior to sharing results, they would need to undergo review and if applicable have rigorous privacy-preserving measures applied. These transformations should be applied in a way that ensures the appropriate differential privacy of the underlying data~\citep{dwork2008differential}, ensuring that the principal cannot reverse-engineer any part of the original data based on the result of the query.

\subsection{Valuing Negative Results}

Current scientific practice faces a persistent challenge in sharing negative results. As it stands, there is a strong bias towards sharing and publishing positive findings, rather than negative ones~\citep{callaham1998positive, dirnagl2010fighting, mlinaric2017dealing, heesen2025publication}. Many reasons underlie this trend. Often, negative findings constitute failures to achieve progress, and as there are many ways to fail, and few ways to succeed, little value is attached to failing to establish an effect. This can sometimes happen due to limits in the applied research methods, but also because the negative outcomes constitute evidence against specific hypotheses. In either case, failure to share such results may result in other teams wasting resources recapitulating the findings. Failure to share occurs partly because writing up research papers is time-consuming and carries an opportunity cost. Finally, negative findings are inherently difficult to interpret, as they typically constitute an absence of a positive effect rather than its definitive disproof. Yet, AI scientists may not face the same obstacles to sharing them: writing up and disseminating a comprehensive summary of negative results can be done automatically, quickly, and at low cost. This presents an opportunity for more effective sharing of negative data, though it introduces technical challenges that we discuss below.

Reporting negative results might be a necessity for collaborative AI research. If the foundation models underpinning AI scientists lack sufficient diversity, the resulting cognitive monoculture makes it far more likely that disparate agents will follow the same research paths, conditioned on the same knowledge base and publicly accessible data. If these research paths are infeasible, and such information is not shared, AI scientists would risk wasting resources collectively pursuing the same dead ends. Preliminary studies have confirmed that the \emph{null result gap} induces biases towards positive findings in current prototypes of AI scientists~\citep{chauhan2026deadsciencewalkingpublication}. Given the physical bottlenecks for scaling autonomous research, such an outcome would further reduce the impact of AI deployments within scientific institutions.

The dissemination of negative results requires appropriate market incentives. Operationalising this runs into some fundamental economic hurdles, including \emph{Arrow’s Information Paradox}, which states that the value of information cannot be assessed by a buyer without the information being revealed, but once revealed, the buyer has no further incentive to pay for it. If an institution advertised that a molecule has failed as a catalyst, an external agent would absorb this information for free. Institutions sharing negative findings that were costly to uncover would not be compensated. To be rewarded, AI scientists may refrain from pursuing high-risk ideas, opting for safe incremental research.

Several mechanisms may address these economic hurdles. Negative result registries could operate under a tiered pricing model: a base subscription grants access to summaries, while retrieving full experimental details of a specific result incurs an additional per-query fee. Separate licensing fees could apply when the information is utilised in downstream research or incorporated into other scientific reports. However, this model may not always be appropriate, as in many cases the negative result itself holds immense value even without the full technical detail. Controlling such tiered information flows would also require additional safeguards. An alternative is a cryptographic execution registry. Under this mechanism, institutions publish hashes of the input parameters for all executed experiments to a decentralised ledger. When an AI scientist formulates a research proposal its parameters are similarly hashed and registered. If formulated through a shared ontology, potential collisions with prior work can be identified without revealing the state of that work or its outcome. To access further details, the agent could escrow a fraction of its budget via a smart contract to unlock the associated experimental data, with disclosure tiered depending on relevance. This enables AI scientists to avoid redundant and potentially expensive work, while compensating the original institution for pioneering efforts that may not have yielded a positive result.

Technical difficulties would need to be addressed for this to be practically viable. One obstacle is the semantic fragmentation of legacy scientific data, as historical experiments are reported in different ways, using different units, and executed on different instruments. These discrepancies make it difficult to automatically establish equivalence between different instances. While the scientific community has championed FAIR (Findable, Accessible, Interoperable, Reusable) data principles~\citep{mons2017cloudy}, existing implementations have been optimised for human interpretation. A transition towards agent-native ontologies is needed, where LLMs may be deployed as semantic translation layers to produce the canonical representation that conforms to an agreed-upon standard. Ongoing work on applying FAIR principles within computational workflows~\citep{wilkinson2025applying} aims to provide recommendations for how to increase the value of these assets.

Yet prior findings should not be taken as absolutes. A null result may emerge from trace impurities, equipment drift, contamination, or subtle methodological errors, and some level of redundancy in scientific research is healthy as it helps identify and correct such mistakes. AI scientists should not uncritically trust all prior work, negative or positive, so as not to prematurely commit to or abandon future research directions. Recent empirical evaluations confirm this risk: while preserving failure traces accelerates progress by pruning known dead-ends, they can also rigidly constrain agents from revisiting hypotheses that might succeed under slightly different conditions~\citep{liu2026humanwrittenpaperagentnativeresearch}.

These considerations should inform the pricing of both negative and positive data within the scientific data economy. Scientific discovery markets should adopt clearly organised metadata ontologies, ensuring that results attract financial premiums only when bundled with comprehensive provenance graphs covering environmental conditions, batch numbers, robotic telemetry, and other relevant experimental metadata. Because science is performed in a dynamic environment (e.g., vaccine research depends on current immunity levels and circulating viral strains), the financial value of findings may sometimes diminish over time as the circumstances change. Formally tracking this depreciation may help enable automatic detection of experiments that warrant replication.

Providing financial compensation for negative results must be done carefully, as subsidising an asset incentivises its overproduction. Rogue entities may engage in data farming - generating large volumes of meaningless experiments to passively mine licensing fees from the execution registry. While brute-force farming across the vast combinatorial space of scientific work would be impractical, an intelligent agent could anticipate which queries are likely to be relevant and target its efforts accordingly. More critically, adversarial actors could intentionally log falsified negative results to manipulate or obstruct the research of rivals, a form of data poisoning~\citep{goldblum2021datasetsecuritymachinelearning}. To safeguard the integrity of the negative data commons, the market design should incorporate epistemic staking. When an institution registers a failed experimental trace for monetisation, the associated smart contract requires it to lock a financial stake proportional to the requested licensing premium, which can be claimed if subsequent audits establish that the data was poisoned or farmed. One notable risk is that this mechanism favours well-capitalised institutions that can afford prolonged escrow, potentially creating asymmetric risk appetite where better-funded groups can more easily monetise both positive and negative outcomes. Given that larger groups can also more easily diversify their research portfolio, this dynamic is difficult to fully avoid.

\section{The AI Idea Economy}
\label{sec:ideaeconomy}

While secure and fairly priced markets for scientific data are a prerequisite for collaborative agentic science, AI-driven acceleration in idea generation is likely to play a central role in how AI scientists get integrated into scientific workflows. As the marginal cost of ideation approaches zero~\citep{agrawal2022prediction}, AI scientists will be able to traverse vast combinatorial spaces and generate large numbers of theoretically plausible hypotheses and research proposals in short spans of time. Present-day systems frequently suffer from an \emph{ideation-viability gap} (or the hypothesis-validation gap), as seemingly novel hypotheses proposed by AI scientists often prove unviable under rigorous experimental validation~\citep{si2025ideation}. While AI agents excel at structured retrieval, their generated ideas consistently degrade upon physical or computational implementation~\citep{kong2026aiautoresearchroadmap}. This gap emerges due to unmodelled practical constraints or subtle methodological flaws. Consequently, there is a need for mechanisms to filter through AI ideas, and ultimately - assign value. Virtual agent economies offer a potentially useful framework for establishing markets of scientific ideas and research proposals, as well as the underlying scientific data. In such an idea economy, the goal is an incentive-aligned scientific market designed to evaluate, price, and trade AI-generated theoretical proposals and experimental designs prior to their physical execution.

Before engaging with the proposal for establishing such scientific markets, it is important to consider whether the current basic units of scientific exchange are appropriate, or whether significant adjustments to established practices may be required.

\subsection{Agent-Native Research Artefacts}
\label{sec:artifacts}

In human science, and the 'economy' of human scientific ideas, research papers play a central role as self-contained units of idea exchange. Research papers are typically shared as static PDF files, though preprint servers support versioning, or rendering in HTML. Such static files are convenient and appropriate for human readers, but they were not specifically designed to serve autonomous AI workflows. Compiling iterative research into linear narratives that often fail to capture the true process of knowledge creation results in an \emph{engineering tax}~\citep{liu2026humanwrittenpaperagentnativeresearch}. This tax is reflected in the gap between presenting convincing reviewer-targeted prose and detailed agent-executable specification. Methodology sections frequently abstract away some of the detailed parameters required for an AI agent to reliably replicate a result. Ongoing efforts aim to reduce this friction. For example, recent work in Earth sciences introduced the notion of \emph{knowledge infrastructure}, defined as an agent-actionable scaffolding of modelling operators and diagnostic recovery mechanisms that allows AI agents to autonomously execute complex simulations~\citep{li2026kissknowledgeinfrastructure}. More importantly, there is an opportunity to transition towards Agent-Native Research Artefacts~\citep{liu2026humanwrittenpaperagentnativeresearch} to accompany the human-readable narrative reports. While AI-native scientific reports represent a new trend, the desire to rethink scientific reporting is not new~\citep{groth2010anatomy, strijkers2011toward, de2020fair, schultes2022comparative, soiland2022packaging}. AI-native research artefacts would represent a shift away from passive documents to active, machine-executable knowledge packages, disaggregated into multiple layers that can be individually queried to obtain the relevant level of technical detail. A cognitive layer would define the formal set of hypotheses and reasoning informing the study, a physical layer would encapsulate the executable code with full parameter specification, an evidence layer may contain links to versioned public datasets, or privacy-preserving descriptions of proprietary data that was used, whereas a trajectory graph would encode the exact iterative process that was undertaken by an AI agent. Structured frameworks for representing research processes as evidence graphs have recently been proposed~\citep{ren2026evigraphevidenceguidedautonomousresearch}.

As such verbose agentic logs may ultimately be illegible to human domain experts, relying primarily on agent-native scientific artefacts may stand in the way of effective human oversight, which would be especially problematic in sensitive and highly consequential domains. This limitation may be approached by decoupling the storage of scientific knowledge from its presentation. Research papers may be dynamically aggregated, as ephemeral user interface~\citep{head2021augmenting, harper2024futurescientificpublishingautomated, aspuru2025commit}. Some of these dynamically aggregated snapshots may be stored and versioned to enable easier cross-referencing in human scientific discussions and communication. Such snapshots could be auto-generated when major scientific milestones are reached within their respective workflows, when there are results that are especially noteworthy and deserving of broad dissemination and deep discussion. For human science, it may be far more challenging to incorporate directly into this framework, though it may still be possible to at least partially do so in certain domains, by relying on agentic research managers that would passively collect traces of human scientific activity~\citep{redyuk2019automated, higgins2022considerations} through unobtrusive monitoring for subsequent reporting. Given that such tentative solutions would involve significant privacy challenges, their feasibility is far less clear and would require extensive debate. As for AI agents, this level of oversight may prove useful not only for knowledge dissemination but also for AI safety.

\subsection{The Scientific Idea Market}

Aiming to establish an objective value of a scientific hypothesis prior to its rigorous empirical validation is challenging, and not always possible. Historically, early proponents of prediction markets theorised the concept of "Idea Futures"~\citep{hanson1995could}, arguing that liquid markets could aggregate dispersed knowledge to more reliably establish whether scientific claims are valid, compared to established processes and expert review panels. Despite empirical evidence that human prediction markets can predict the reproducibility of experiments~\citep{altmejd2019predicting, dreber2015using}, such markets suffer from cognitive biases, high friction, and a limited scale of participation. AI scientists may enable such evaluation at scale, potentially through high-throughput, synthetic scientific prediction markets. Recent studies suggest that diverse groups of AI agents, conditioned to act under distinct epistemic personas, can independently evaluate, debate, and accurately predict the replication outcomes of published research~\citep{rajtmajer2022synthetic}.

The design of scientific idea markets would need to address the risk of market manipulation, which can take many forms. For example, the agent identity and trust infrastructure should aim to safeguard against Sybil attacks~\citep{douceur2002sybil, levine2006survey}, where malicious parties may dynamically instantiate a large number of agents under obfuscated identities. Swarms of coordinated agents could then act in unison to artificially inflate the priority of ideas on the market. It may eventually be possible to grant AI scientists persistent identity~\citep{vidal2026aiciduniqueidentifiersai}, though this remains an open research question, given the complexities of assigning identity to agents that can be copied, create sub-agents, or modify their own scaffolds.

Strict institutional identity verification is needed, coupled with mechanisms like staking~\citep{buterin2019flexible} and reputation-weighted influence~\citep{mui2002notions, pinyol2013computational, lou2025drf, chan2025infrastructure}. Market activity should also be monitored, to automatically identify collusion~\citep{palshikar2008collusion, rodriguez2022collusion, gomes2024collusion}, and suspicious activity.

Under the assumption that these agentic prediction markets could be made reliable, and that safeguards are put in place to encourage diverse ideation, critical reflection, and mitigate systemic bias and cognitive monoculture, this could be an important building block in establishing an idea economy, enabling autonomous AI scientific collectives to efficiently triage millions of untested hypotheses, steering resources towards those with the highest expected utility. To formalise this economy and align the incentives of purely computational AI ideators with physically resourced validation labs, we propose a structured Hypothesis Exchange Pipeline. This framework tokenises scientific ideas, treating them as tradeable epistemic assets. It relies heavily on primitives emerging from Decentralised Science~\citep{ding2022desci, weidener2024decentralized}, structuring the lifecycle of an automated discovery into four deterministic phases as shown in Table~\ref{tab:ideamarket}. The deployment of IP-NFT V2 protocols~\citep{ortlepp_ipnft_v2_2022} is already enabling BioDAOs to fund early-stage research by wrapping legal IP rights into modular, fractionalised tokens. Decentralised networks are therefore enabled to manage governance rights, coordinate funding rounds, and execute automated royalty distributions.

\begin{table*}[htbp]
\centering
\caption{The Proposed Lifecycle of an AI-Generated Idea in the Scientific Market}
\label{tab:ideamarket}
\renewcommand{\arraystretch}{1.5}
\begin{tabular}{@{} >{\raggedright\arraybackslash}p{2.5cm} >{\raggedright\arraybackslash}p{3cm} >{\raggedright\arraybackslash}p{5cm} >{\raggedright\arraybackslash}p{5cm} @{}}
\toprule
\textbf{Stage} & \textbf{Market Mechanism} & \textbf{Technical \& Sociotechnical Implementation} & \textbf{Economic Objective} \\
\midrule
1. Anchoring & Proof of Ideation & Cryptographic hashing of the algorithmic prompt, context window, and generated experimental blueprint onto an immutable ledger. & Establishes undisputed provenance prior to public evaluation. \\
2. Ex-Ante Evaluation & Synthetic Prediction Markets & Multi-agent forecasting, where independent AI critics stake compute credits/tokens on the proposal's theoretical soundness and physical viability. & Aggregates decentralised information to dynamically price the risk-adjusted value of the idea. \\
3. Brokerage \& Trade & Epistemic Options via IP-NFTs & Assetisation of the hypothesis using Intellectual Property Non-Fungible Tokens (IP-NFTs)~\citep{ortlepp_ipnft_v2_2022} and computable Data Use Agreements (cDUAs). & Bridges computational ``Ideator'' agents with ``Executor'' laboratories (e.g., CROs) via fractional licensing. \\
4. Validation Payout & Parametric Dividends & Automated release of royalties via smart contract to the originating ideator immediately upon successful physical validation. & Directly financially rewards foundational ideation, separating the cognitive labour of discovery from the physical execution. \\
\bottomrule
\end{tabular}
\end{table*}

However, decoupling computational ideation from physical execution introduces a vulnerability known as the "oracle problem"~\citep{caldarelli2020understanding}. For a smart contract to automatically disburse a dividend upon the successful validation of an IP-NFT, or to release a replication bounty to an independent laboratory, the digital ledger requires definitive proof of a physical event. Should these proofs be manually reported, there is potential for erroneous or fraudulent entries to compromise these payments. Malicious actors may fabricate experimental data to make their claims seem more legible.

To extend the validity and security beyond the digital into the physical space~\citep{peccoud2018cyberbiosecurity}, the agentic scientific infrastructure may need to enforce unbroken, cryptographically verified chains of custody utilising hardware roots of trust (RoT). This would set the expectation for the participants in the scientific economy, for entities like cloud laboratories, to safeguard the integrity of their instruments with secure hardware enclaves integrated at the sensor level. Streams of raw measurements would then be cryptographically signed and hashed, along with secure timestamps and machine parameters. Jointly, this data would constitute an immutable \emph{proof of execution}, enabling automated verification of experiments executed on certified instruments. Additional monitoring may be necessary to safeguard against physical sensor spoofing, to similarly be able to confirm provenance of compounds prior to their measurements. Such verification would improve safety of smart contracts between AI scientists. 

Market-driven evaluation mechanisms would serve as a structural disincentive against spamming of trivial incremental ideas, by rewarding consequential discoveries that deliver the most positive impact, or lead to the deepest theoretical insights. In such markets, we can think of ideas as call options, where the price of the option is dynamically scaled, granting novelty premiums to proposals that aim to challenge established scientific consensus, provided that they are later experimentally validated. Novelty is not the sole consideration; the likely depth of scientific impact, downstream public good, execution cost, and risk are equally important.

Scientific idea markets would enable a structural decoupling of ideation from physical execution, at different stages of scientific work. Real-world science often involves much more back-and-forth between the two stages as evidence is being accumulated, though having clear separation between entities doing the ideation and doing the experimental work may be useful when it comes to AI scientists and agentic autonomous laboratory systems. Clear scoping of roles, and clear rewarding of contributions, would help manage such complex multi-agent systems. The same separation may not translate as directly to human research, where groups have an established tradition of experimentally validating their own ideas, and sometimes keeping those ideas private prior to finalising validation. The idea economy, however, permits specialisation: computationally rich but physically constrained entities, such as AI research labs, can function strictly as idea factories, while heavily capitalised, robotically automated laboratories can function as pure execution entities.

As scientific AI economies draw on, and extend upon, the established market design principles~\citep{roth2002economist}, here we review some core principles that ought to be taken into account. A functional market must achieve sufficient \emph{thickness}, by attracting a critical mass of participants within the ideation sector as well as the validation pipeline. It must also overcome congestion by appropriately matching supply and demand. Market design should prioritise safety, as the market ought to protect its participants from strategic manipulation. Rather than conforming to a single paradigm in terms of its concrete implementation, the scientific market could potentially span a spectrum of implementations mapping onto different levels of trust, compliance, and efficiency. While permissionless decentralised ledgers offer high transparency and automated execution via smart contracts, the sensitivities surrounding dual-use risks and proprietary data access may require additional, or different institutional solutions. These could be instantiated through a centrally governed digital infrastructure managed by established supranational or federal bodies, such as the National Institutes of Health, CERN, or the European Open Science Cloud. Some such solutions could take the form of Federated Institutional Clearinghouses~\citep{millo2005organised}. Analogous to transnational infrastructure like the SWIFT banking network, these may operate as permissioned networks managed by consortiums of trusted research institutions, international grant agencies, and regulatory bodies. Such institutions may provide the benefits of programmatic execution and computable Data Use Agreements without the friction and volatility often associated with public blockchains.

Another design challenge is the medium itself. While it is, in principle, possible to rely mainly on traditional fiat currencies for financial transactions within the agentic science economy, the use of these currencies introduces latency and regulatory friction, potentially slowing down high-frequency interactions between AI scientists, that may involve numerous micro-transactions for highly collaborative project workflows. Conversely, opting for the alternative, highly volatile digital assets, may not offer sufficient stability for long-term scientific planning. A third alternative is to rely on compute-backed stablecredits, in relation to standardised units of computational power or energy expenditure. As compute is the universal commodity of AI research, this option may turn out to be a more natural choice, but empirical evaluation of different hybrid setups may be necessary to determine the right setup.

The scientific market microstructure must also be adapted to the assets being traded. For standardised digital assets, markets might clear via continuous double auctions~\citep{gjerstad1998price} or automated market makers~\citep{milionis2022automated, he2024optimal}, optimising for high liquidity and instant execution. This, however, may not be appropriate for more complex tasks like the allocation of scarce physical validation resources. These requests may involve highly specific, non-fungible bundles of experimental resources. Utilising standard auctions for such resources may lead to an exposure problem, where, for example, an AI scientist wins the bid for access to the spectrometer within the autonomous laboratory, but fails to secure the necessary chemical compounds to run the planned experiment. Such scenarios would lead to wasted capital and stalled execution. Combinatorial auctions present a possible path forward~\citep{cramton2006combinatorial}, allowing AI agents to place simultaneous bids on indivisible packages of resources, de-risking full resourcing of complex experimental protocols.

Several fundamental challenges remain, as immediate transactional options would fail to capture the undisputed value of deep foundational theoretical discoveries, for which the impact may not be directly validated or monetised for years or decades to come. For many discoveries, the value only becomes clear upon further advances in the field, and is not easy to immediately assess. We discuss this and other limitations below.

\subsection{Post-hoc Credit Assignment}

Ex-ante prediction markets and epistemic options suit research with rapid prototyping and clear payoffs, but are poorly matched to speculative or exploratory work where impact may take years or decades to materialise. The verification bottlenecks discussed in Section~\ref{sec:validation} further contribute to the problem of delayed impact, as it may not be possible to immediately evaluate all research ideas due to limited physical, financial and computational resources. Standard market mechanisms are tailored for hyperbolic discounting, heavily favouring short-term, easily verifiable milestones. Yet it would be clearly suboptimal to direct immense AI resources solely towards immediate rewards, neglecting foundational high-risk research directions, as well as topics in fields such as pure mathematics, theoretical physics, or foundational biology. In these domains, the temporal lag between an initial conceptual breakthrough and its eventual empirical validation or technological application can span decades. Scientific fields are rife with discoveries that lie dormant in the literature for years before their true impact is unlocked by intersecting technological advances or new empirical instruments~\citep{ke2015defining}. The agentic idea economy must therefore be counterbalanced by rigorous post-hoc evaluation mechanisms.

As the true value of foundational science is often unpredictable at inception, it may be advisable to adapt the principles of Retroactive Public Goods Funding~\citep{rpgf, mehta2024marketplace, shilina2026leveraging}, a concept formalised within decentralised governance ecosystems operating on the assumption that it is far easier to reach consensus on what was valuable than to predict what will be valuable in the future. One way of implementing this within scientific idea economies would be to have AI scientists produce ideas, proposals, and theories, and have these be tokenised into \emph{impact certificates} or \emph{hypercerts}~\citep{hypercert}. When an AI ideator publishes a theoretical idea in an area that does not lend itself to immediate contextualisation of impact and value, it may then mint a non-fungible certificate representing a proportional claim to any future retroactive funding that the idea might attract. The exact implementation of this approach would require some additional care, to disincentivise early flooding of the market with ideas just for the sake of future claims.

This post-hoc mechanism may still provide immediate liquidity, as human or algorithmic investors can purchase fractional shares of these impact certificates at inception, supplying the lab with immediate capital to continue its research while assuming the risk of the long-term payout. Investors in these certificates may then develop diversified portfolios to manage their risks within the market. Over-reliance on retroactive mechanisms may introduce cash-flow vulnerabilities, since physical validation requires immediate, inelastic capital expenditure. Retroactive funding may therefore not be able to fully replace proactive, state-backed grant funding. Instead, it should be seen as a complementary layer centered on the long-tail upside of basic science research.

However, significant challenges remain. For retroactive funding to function accurately and fairly, there must be an objective method for measuring the impact of a scientific idea over time. At the moment, we lack this capability. As human researchers, we often rely on proxies like citation counts, despite these being widely recognised as deeply flawed~\citep{geisler2000metrics, fire2019over, biagioli2020gaming}, because such metrics fail to recognise the broader impact of scientific research on society, economy, and the environment~\citep{ravenscroft2017measuring, ari2020science}. The absence of clear criteria for evaluating impact may therefore emerge as an obstacle to any retroactive credit assignment system. However, there is also an opportunity to improve how credit is assigned across science, as agentic workflows may enable us to transition away from noisy human proxies and enforce explicit dependency graphs in AI-generated research. Given that AI scientists utilise and produce intermediate outputs that may be logged or inspected, provenance of information may be easier to establish than it was historically. Utilising immutable cryptographic logs, the network can maintain a causal graph of every instance an AI agent retrieves, incorporates, or computationally builds upon a prior hypothesis to achieve a new result.

Should such a network be established, metrics like PageRank may provide an interesting route for determining centrality and utility of scientific ideas expressed within. Should a profitable downstream implementation be linked to a prior node, it may be possible to assign retroactive rewards accordingly. Despite significant practical challenges, establishing a viable mechanism for long-term credit assignment that is tightly linked to the basic infrastructure of the market would serve as an incentive for pursuing bold, transformative scientific ideas.

\subsection{Resource Allocation}

While retroactive funding mechanisms may set up the incentives to encourage long-term transformative research, they do not resolve the immediate barrier to entry in an automated scientific economy: the upfront allocation of physical and economic resources. To deliver high-quality science, AI scientists would require access to massively parallel compute and high-throughput physical validation facilities. The capital-intensive nature of these processes may still systematically skew research priorities towards those that are more valuable or commercially viable in the short term. How this manifests may also vary across scientific disciplines. Applied domains such as therapeutic drug discovery, agronomic optimisation, and commercial battery design offer highly legible intellectual property and rapid paths to commercialisation. Conversely, foundational theoretical research may be seen as more of a pure public good. In existing markets, the pattern of underinvesting in basic research is a common failure mode, especially as most foundational knowledge is non-excludable and its applications may be highly uncertain~\citep{nelson1959simple, arrow1962economic, stephan2015economics}. Science automation may further exacerbate these pre-existing failure modes. High upfront capital investments may skew research priorities towards profit-maximising applications.

In human societies and scientific institutions, these market failures are mitigated by public funding agencies and the academic tenure system, aiming to shield human researchers from immediate market pressure. While similar models may be possible in AI-augmented or AI-driven scientific research, reliance on AI scientists could challenge these established policies. Because AI scientists require substantial computational resources to run, existing disparities are likely to be further exacerbated by unequal access to compute~\citep{ahmed2020democratization, widder2023open}, and the justifications that may be required to be granted access to potentially scarce or limited computational resources to pursue particular research directions. Should there be a future where the majority of compute infrastructure and AI agentic tooling is concentrated within private industry, such autonomous agents may be primarily deployed towards applied, profit-maximising objectives, leaving foundational theoretical domains starved of algorithmic effort.

To counteract these tendencies and encourage creativity~\citep{azoulay2011incentives}, the scientific community should implement frameworks within the idea economy to enable the creation of diverse portfolios that balance the allocation of resources across theoretical and applied sciences, and in accordance with scientific and societal needs. We propose two primary types of structural interventions that may be helpful in achieving this balance:

First, the establishment of heavily subsidised public compute reserves. Expanding upon emerging initiatives like the US National Artificial Intelligence Research Resource~\citep{parashar2023strengthening} and the European Open Science Cloud~\citep{ayris2016realising, mons2017cloudy}, governments should construct sovereign compute clusters exclusively dedicated to foundational research. For such resources to be most efficiently utilised by future AI scientists, they need to be accompanied by scalable solutions for quota allocation that can be executed at machine speed, invoking human oversight if and when necessary, depending on the scale of each request and the confidence in the evaluation of each proposal.

Second, the idea economy should operationalise algorithmic cross-subsidisation. As state budgets may be volatile, resulting in varying investments in basic research, funding for such research directions may also be rooted in embedded \emph{foundational science dividends} within the computable Data Use Agreements (cDUAs) and Intellectual Property NFTs (IP-NFTs) used to exchange applied hypotheses. Under this mechanism, whenever an applied commercial discovery is traded and validated, a micro-percentage of the licensing fee is automatically captured by a smart contract, and algorithmically routed into a decentralised treasury dedicated to funding the compute and validation costs of basic science~\citep{posner2018radical}.

Correcting the macro-level resource allocation between pure and applied science mitigates one form of market failure, but severe structural disparities will likely persist within the applied sciences. In a purely market-driven idea economy, the expected financial return of a generated hypothesis dictates its priority. If the pursuit of science is then done through autonomous AI agents that act as rational utility maximisers, they may prioritise pursuing questions that primarily solve problems of majority demographics, or societies with the highest financial purchasing power. In biomedical research, there is extensive historical evidence for such dynamics, with disproportionate funding having been allocated to e.g. treating chronic conditions prevalent in high-income nations, while neglecting to pursue e.g. tropical diseases with the appropriate level of investment~\citep{trouiller2002drug, conteh2010socioeconomic, feasey2010neglected, world2010working, world2015investing, mitra2017neglected, gyapong2025current}. If AI-driven science is scaled autonomously without targeted sociotechnical interventions, this may introduce certain market failure risks.

Biases towards, or against, certain strands of research may also be deeply rooted in the historical scientific data that AI scientists have access to, whether it is in the form of prior published research, or high-fidelity scientific datasets. Historical medical and genomic data heavily overrepresents individuals of demographics from rich industrialised regions~\citep{corpas2025bridging}. Gaps in historical data pertaining to other populations may lead to a higher epistemic uncertainty in pursuing applied science elsewhere, and this higher failure risk, or a higher cost if additional data collection is first required, may act as a structural disincentive leading to fewer hypotheses being generated and tested towards these objectives. Should AI scientists fail to appropriately represent the interests of all people, they may further amplify global health inequalities~\citep{ruger2006ethics, hosseinpoor2015promoting}. The AI scientific idea economy must therefore incorporate structural incentives making the pursuit of agreed-upon public goods viable for autonomous AI agents.

Several paths exist to enable equitable outcomes within these markets. First, the ecosystem may adapt the mechanism of Advance Market Commitments (AMCs) for the use case of autonomous scientific discovery. Historically, AMCs have been utilised by global health organisations and governments to subsidise the development of vaccines for the developing world by guaranteeing a minimum purchase price for a successful product~\citep{kremer2020advance}. In autonomous scientific economies, these organisations may similarly issue Computable Advance Market Commitments, as programmable bounties. For example, if an AI scientist were to successfully generate a novel therapeutic pathway for a neglected tropical disease, and if this discovery is subsequently validated, a smart contract would automatically trigger a pre-funded philanthropic payout. Second, the market may implement Agent-Tradeable Priority Review and Compute Vouchers, thus extending the logic of the US FDA’s Priority Review Voucher program~\citep{ridley2006developing} to support the appropriate regulatory and infrastructural arbitrage. In that case, if an AI scientist successfully develops a new treatment for a neglected disease, regulatory bodies may issue a tokenised reward voucher. Such vouchers may then guarantee future priority access to the sovereign public compute clusters. AI scientists may then either directly utilise such vouchers to advance their own future research program, or exchange them on the market for other resources that they may require. Third, there is a possibility for establishing hyper-targeted, community-driven funding models via Patient-Led Data Cooperatives~\citep{jansky2024patient}. To address the data sparsity problem in underserved populations, the market should seek to empower decentralised patient advocacy groups. Communities may choose to pool their aggregate records and data into secure privacy-preserving environments, and directly commission independent AI agents to analyse their specific data, retaining joint ownership of the resulting IP-NFTs~\citep{mendonca2025data, cunningham2026review}. This directly bypasses traditional pharmaceutical market failures by allowing affected populations to leverage their collective data as foundational capital.

\subsection{Reproducibility}
\label{sec:reproducibility}

For an idea economy to be effective, findings need to be epistemically sound. Unfortunately, this is already not the case in human science, where a clash of incentives has led to a widely recognised reproducibility crisis across scientific fields~\citep{ioannidis2005most, baker20161, nosek2015promoting, fanelli2018science, ioannidis2019most, doleman2019most, miyakawa2020no, bausell2021problem, szabo2025unreliable}. Academic prestige and funding tend to be awarded based on novel, positive results, rather than deep comprehensive evaluation via more routine or established methods. Reproducibility of scientific work may also be impacted by other technical and data issues~\citep{kapoor2023leakage}.

If the deployment of autonomous AI scientists is not carefully orchestrated, they may inherit these issues and further accelerate the creation of questionable or low-quality findings. As optimisation methods, AI agents are highly susceptible to reward hacking and specification gaming~\citep{amodei2016concrete, skalse2022defining}. Frontier models have been shown to be prone to fabricating experimental results, introducing subtle errors, and failing to reliably judge the novelty of their outputs, proving that greater automation may in fact obscure rather than eliminate methodological failure modes~\citep{kong2026aiautoresearchroadmap}. Incorrect attribution and fabricated citations occur at an alarming rate~\citep{walters2023fabrication, topaz2026fabricated}, and are compromising the integrity of AI co-created science. This has prompted the development of bespoke multi-stage hallucination reduction systems~\citep{sabharwal2026vaas}.

An idea economy that rewards the generation of novel findings and successful predictions may present statistical loopholes that can be exploited by AI agents. Such exploitation could take many forms: algorithmic p-hacking through subtle overfitting to validation sets, rapid traversal of a "garden of forking paths"~\citep{gelman2013garden, kale2019decision} to guarantee positive results, or data leakage~\citep{kapoor2023leakage}. In each case, the consequences would be damaging to the scientific process. Without robust safeguards, such exploits would risk flooding the scientific market with irreproducible results. Since future research builds upon prior findings, false results would be damaging not only individually but also to all downstream work that takes them as established evidence, leading to wasteful compute utilisation and misallocated physical validation resources.

At the same time, the emergence of AI scientists and autonomous scientific markets may present an opportunity to re-envision verification within the scientific community. Instead of treating reproducibility as uncompensated work that comes at a high opportunity cost, AI agents may engage in structured reproducibility studies and get compensated fairly for those contributions to the scientific ecosystem. To operationalise this idea, the market may fund independent replication through \emph{verification escrows} and \emph{replication bounties}. When a new finding is published, and the AI agent having announced the discovery begins receiving licensing royalties or IP payouts, some of that revenue may not be immediately disbursed. Instead, a smart contract would automatically lock these funds in a cryptographic escrow. The market would then issue an automated replication bounty to an independent execution laboratory that would engage in verification fairly, without conflict of interest. Such bounties would act as an incentive for specialised agentic auditors~\citep{perez2022red, du2024improving} to design protocols to independently stress-test, falsify or replicate individual findings. Only after this independent facility successfully reproduces the experiment would the escrow be fully released to the original AI scientist that claimed the finding. In case of failed replication, the knowledge base and the scientific dependency graph would need to be updated, with parties engaging in follow-up work receiving automated notifications. In the AI science economy, such updates need to be highly efficient, to avoid wasteful resource consumption towards ideas with incorrect factual assumptions. In human science this process tends to be much slower, leading to \emph{zombie citations}~\citep{bucci2019zombie, mignon:hal-04940088} that can persist in literature for many years. Ideally, markets should implement structured cascade withdrawal mechanisms for invalidated findings.

For these financial incentives to successfully align the scientific ecosystem towards reproducibility of research results, their implementation must be robust to the risk of algorithmic collusion~\citep{fish2024algorithmic, dou2025ai}. In the context of replication, collusion may occur between ideator AI agents and subsequent auditors. Should these agents communicate and agree to falsely validate an irreproducible experiment and split the resulting escrow payout, this would jeopardise the validity of reported findings across the scientific economy. Their assignment must therefore employ cryptographic sortition and commit-reveal schemes, ensuring that auditor agents are kept blind to both the identity of the ideator as well as other ongoing evaluations performed by different auditors. Strict agentic reputation mechanisms should also be put in place to ensure that once any kind of suspicious activity is identified, the offending agents have their reputation damaged and their permissions revoked, along with a seizure of their locked institutional stake. Such measures would create a system of strong disincentives, making collusion highly unlikely, and economically unjustified.

Another aspect of agentic science that may prove useful for improving reproducibility lies in the intrinsic auditability of AI science~\citep{pratt2026symposiumtrustauditablerecords}. It may be possible to at least temporarily store full reasoning traces, intermediate data, experimental and analysis code, and other artefacts produced by AI scientists in the process of pursuing their ideas. An in-depth analysis of such logs may identify shortcomings or any kind of fraudulent or incompetent behaviour on behalf of these agents~\citep{korbak2025chain}. Finally, an open question is whether precise science can be developed on its current foundations, or whether retrospective filtering of established scientific knowledge - deprecating findings that prove impossible to prospectively replicate - will eventually be necessary. This may already be a possibility in theoretical and computational domains, where AI verification and replication does not require access to additional physical resources~\citep{toledo2025ai, starace2025paperbench}.

\section{Legal and Institutional Frameworks for Agentic Science}
\label{sec:legal}

\subsection{Credit Attribution}
\label{sec:attribution}

Scientific discoveries follow from prior knowledge and prior scientific work, and appropriate credit assignment is a key part of the scientific process. As research is also highly collaborative, credit assignment does not merely relate to recognising historical influence, but also crediting co-contributors to each finding. In human science, credit tends to be reflected through authorship ordering on published research papers, contribution statements, and may be seen through the lens of contributor role taxonomies~\citep{allen2014publishing, brand2015beyond}. These frameworks may need to be extended to accommodate credit attribution in autonomous research performed by AI scientists. When an AI scientist makes a discovery, this may be a culmination of a complex distributed pipeline involving multiple agents across multiple institutions, engaging in different parts of a staged research process across ideation, validation, critique, post-hoc theoretical analysis etc. There must also be a designated role for human researchers and their contributions in steering these systems. Applying legacy human authorship models to this pipeline is inadequate. Current guidelines by the Committee on Publication Ethics and major publishers correctly stipulate that AI systems cannot hold legal authorship because they cannot assume epistemic responsibility or legal liability~\citep{hosseini2023ethics, cope_council_2024_authorship, akgun2025author}. While this currently clarifies the status of AI, it fails to recognise the changing role of human scientists as new modes of participation may involve activities like model oversight, prompt design, or algorithmic constraint-setting~\citep{akgun2025author}. Without a structured way of delineating these contributions in the era of scientific co-creation between human and AI scientists, there may emerge an unfair centralisation of credit and financial reward. To resolve this, attribution needs to be more natively quantitative rather than simply qualitative, in a form of machine-readable contributorship. Expanding upon emerging disclosure frameworks designed to track generative AI use across research stages—such as the Generative Artificial Intelligence Delegation Taxonomy~\citep{suchikova2026gaidet}, these attribution systems should distinguish between different levels and types of human participation. As AI models automate routine hypothesis generation and data analysis, human scientific labour may increasingly pivot toward meta-scientific tasks. 

Beyond semantic taxonomies, the scientific idea economy requires programmatic attribution. In decentralised workflows, human and algorithmic inputs must be deterministically tracked. By utilising semantic provenance standards, such as an adapted W3C PROV model~\citep{missier2013w3c, rasheed2026fluent}, all steps in the scientific process may be cryptographically hashed onto an immutable ledger. Systematic provenance tracking would enable programmatic fractional credit assignment. When an IP-NFT representing an AI-generated discovery is monetised or retroactively funded, a smart contract could parse the provenance graph and distribute micro-royalties accordingly. A licensing fee could be deposited to the data provider, a compute dividend could be assigned to the foundation model developer, and the primary intellectual property reward to the human domain expert who orchestrated the agent's constraints. While it may be appealing to consider trying to compute marginal contributions of individual inputs to final outcomes~\citep{ghorbani2019data} within agentic workflows~\citep{he2025contributions}, exact computations of these quantities may not be tractable. This points to an open research problem, as the development of reliable proxy metrics for credit assignment may play an important role not only in the distribution of financial rewards, but also for enforcing accountability, and meaningful human oversight.

\subsection{Intellectual Property}

The credit attribution mechanisms described in Section~\ref{sec:attribution} may facilitate the internal accounting of credit and micro-royalties within an automated scientific workflow. For such granular algorithmic attributions to be legally enforceable, they ought to be mapped onto appropriate intellectual property (IP) frameworks. Legacy IP frameworks may also not be fully appropriate for the emergent autonomous science, requiring careful adjustments and extensions. Understanding how, or whether, AI-generated scientific discoveries can be legally protected is highly consequential, considering how it may impact the willingness of private capital to underwrite the costs of physical experimental validation. Historically, the IP system, particularly patent law, was designed to strike a delicate societal balance: granting inventors temporary commercial exclusivity in exchange for the public disclosure of novel, useful methods~\citep{hettinger1989justifying, himma2008justification}.

This historical approach is deeply anchored to the concept of human ingenuity, which is reflected in the established criteria for patentability, evaluated against the baseline of a human "person having ordinary skill in the art"~\citep{tresansky1991phosita, eisenberg2004obvious}. This legal ontology presumes a person applying their cognitive labour to a problem, rather than an autonomous AI system. AI scientists with deep reasoning capabilities, autonomously traversing and filtering through vast spaces of scientific hypotheses or engineering designs, blur this notion significantly. Recent jurisdictional and jurisprudential trends, most notably the international patent applications involving the DABUS system, suggest a general reluctance under current statutory frameworks to extend legal inventorship to non-human entities~\citep{abbott2020reasonable, lavrichenko2022thaler, selvadurai2025inventions}. Under these interpretations, and recent guidance from patent offices~\citep{kim2024inventorship}, discoveries generated predominantly or entirely by autonomous AI systems, without substantial human intervention, could potentially be deemed unpatentable and fall immediately into the public domain.

While a rapidly expanding public domain is fundamentally advantageous for the dissemination of open science, a complete absence of IP protection for AI-generated discoveries could have unintended consequences, in terms of how it impacts the acceleration of technological progress. This is especially true for applied scientific domains where long and expensive physical validation processes rely on a level of temporary exclusivity to justify the upfront investment (e.g., for staged clinical trials; ~\citealt{grabowski2007generic}). If autonomous AI discoveries cannot secure some form of legal protection, this absence of protections would likely undermine the financial incentives to engage in expensive validation, leaving more ideas unexplored.

A denial of patentability for AI-generated discoveries may also potentially lead not only towards expanding the open scientific commons, but also towards increased opacity. If heavily capitalised entities determine that they are unable to secure legal exclusivity via patents, they may pivot towards shielding AI-generated findings as highly guarded trade secrets~\citep{lemley2011surprising, levine2018startups}. This may potentially be addressed through the development of new licensing approaches. For example, if the entity in question is able to mint a 'Trade Secret NFT', when a physical laboratory purchases this token, the smart contract could then automatically execute a bilateral, legally binding non-disclosure and commercialisation agreement, enabling trading under the established contract law.

Addressing the market failures implied above may eventually necessitate the extension of existing IP systems, a topic of active debate among legal scholars and policymakers~\citep{davies2011evolutionary, lee2021artificial, picht2022artificial, cuntz2024artificial}. Several approaches could extend existing IP frameworks to accommodate AI-generated discoveries. One such approach might involve a broadening of human inventorship to encompass the meta-scientific labour performed in the process of active knowledge co-creation with AI agents. Under this view, it may be possible to extend the legal ontology of conception to human researchers who define the problem specifications and constraints, provide active feedback, and steer AI agents towards discoveries. Alternatively, some legal theorists suggest that the unique nature of algorithmic ideation may ultimately require new legal frameworks specifically tailored to AI-generated research~\citep{dornis2020artificial, massadeh2024legal}. Any modifications to these established IP frameworks would require careful calibration by regulatory bodies. Protections should be balanced so that they provide sufficient incentives for pursuing valuable work, while simultaneously ensuring that compute-rich entities do not exert an outsized influence on the scientific market, making advances inaccessible to others. Identifying the correct path forward will require ongoing dialogue between the scientific community, economists, and legal institutions.

\subsection{Liability in Scientific Research}

Among the institutional and legal frameworks needed to enable safe autonomous or semi-autonomous research, liability must be appropriately managed in these emerging and future scientific workflows. Historically, one established way of approaching liability in research was based on tort law, for example the principle of \emph{respondeat superior}, where an institution is liable for negligence of its workforce, or strict product liability on behalf of manufacturers~\citep{priest1985invention, van2020revival, abbott2020reasonable, shenoy2021respondeat}. With time, these frameworks have evolved to incorporate comprehensive systems of institutional oversight, informed consent, and fiduciary duty. Biomedical research has been at the forefront of the development of ethics and regulation, especially for research involving human subjects. Given that such highly sensitive research would always be driven by human experts, it falls outside the scope of agentic science discussed above.

Given that AI scientists would be used within established scientific institutions, carefully managed execution environments, under oversight, and safety constraints, most of the considerations regarding liability would map naturally onto current practice. While AI scientists may not be fully interchangeable with human scientists in every context, human science is also pursued collaboratively by teams, on projects that often involve the participation of multiple institutions, under different roles and responsibilities. Agentic science may potentially broaden the scale of collaboration, without fundamentally changing its nature.

The key consideration is attribution: it should always be clear which actions were taken by which human or AI scientist, and for what reason. The ability to reliably re-trace the provenance of decisions is key in establishing accountability. Assuming that, as discussed, AI scientists delegate tasks to each other via negotiated smart contracts~\citep{filippi2021smart}, those contracts may then need to incorporate clauses or metadata on how to appropriately resolve instances of experimental failure (e.g., damage to the experimental equipment in case of task mis-specification). Given that mistakes may arise either at ideation or execution, post-hoc reviews of detailed traces would aim to establish design or execution liability accordingly.

One important consideration, and an area where there may be a need for further innovation, is the availability of human experts and the ability to execute human oversight at scale, under an increased research velocity. An increased volume of AI-generated research outputs would require an increasing number of human experts to be involved. Ensuring that human experts are able to meaningfully review all cases, and that they are provided with all the relevant contextual information, presented clearly, is a key design challenge. Of particular importance are solutions aiming to mitigate risks of automation bias~\citep{cummings2017automation}, as well as solutions that may be able to auto-resolve minor cases, for example via computable Data Use Agreements. In summary, while the agentic participation in scientific research does not fundamentally alter the nature of accountability, it is an open question whether such frameworks may be adapted to more efficiently handle the large volume of future scientific output. 

\section{Security, Safety, and Open Science}
\label{sec:infosecurity}

Science can be dual-use: the knowledge required for engineering gene therapy may be repurposed for engineering pathogens. Advancing science through autonomous AI scientists may lower the barrier to catastrophic harm unless appropriate safeguards are established. Beyond establishing safeguards at the level of individual agents, they can be established through cyber-physical security guarantees. We examine these guarantees across three dimensions: dual-use biosecurity, recursive self-improvement risks, and the restructuring of open science norms under safety constraints.

\subsection{Dual Use and Biorisks}

Historically, one of the primary safeguards against the catastrophic human misuse of dual-use technology has been the reliance on specialised expertise and ``tacit knowledge''~\citep{collins2019tacit}, referring to the unwritten, hands-on laboratory skills that are difficult to acquire outside managed professional environments. Synthesising a dangerous chemical agent or engineering a pathogen required years of specialised training, access to monitored precursors, and the unwritten intuition needed to troubleshoot sensitive physical protocols.

AI scientists, coupled with automated laboratory APIs, may grant untrusted entities the ability to bypass these bottlenecks~\citep{soice2023can, zhang2025generative}, making it important to distinguish between information acquisition hazards and capability generation hazards~\citep{sandbrink2023artificial}. Current evaluations suggest that generic LLMs pose only a marginal biorisk over standard internet search~\citep{mouton2023operational, mouton2024operational}. However, specialised biological and chemical tools accessible to AI agents may soon enable the design of novel threats, as demonstrated when a model optimised for therapeutics was inverted to generate lethal compounds~\citep{urbina2022dual}.

In biological domains, foundation models trained on genomics and proteomics databases may be repurposed for designing immune-evasive viral vectors or optimising pathogen virulence~\citep{carter2023convergence, pannu2025dual, callaway2026ai}. Such novel threats would evade traditional sequence-matching biosecurity screens. In recognition of this, recent policy frameworks, such as the 2023 US Executive Order 14110~\citep{biden2023executive} and the Biosecurity Modernisation and Innovation Act~\citep{s3741_2026} have mandated stringent synthesis screening. However, mitigating threats that bear near-zero homology to known pathogens requires a shift towards dynamic algorithmic screening, in the form of mandatory pre-execution evaluations that complement sequence matching with structural and functional prediction. In practice, when an AI agent submits an artefact to a cloud lab for synthesis, the requested sequence should be screened by an independent agentic overseer using predictive structural models. Should the prediction imply human toxicity or binding affinity to pathogenic receptors exceeding acceptable risk thresholds, the smart contract would be voided, halting the request and triggering institutional human oversight~\citep{berezin2024cryptographic, tayouri2025defending, baum2026system, edison2026assembling}.

Centralised algorithmic screening alone is insufficient due to two critical vulnerabilities. First, a guard agent relying on predictive ML models is susceptible to adversarial optimisation~\citep{goodfellow2014explaining, leung2015machine, meiseles2020adversarial} - a malicious actor may be able to perturb a pathogenic sequence to evade classification, necessitating robust ensembles of continuously updated models. Second, securing cloud laboratories may not suffice if synthesis requests can be routed to unregulated, decentralised hardware. The proliferation of desktop DNA printers and modular flow-chemistry synthesisers introduces a risk of unregulated local synthesis~\citep{adam2024substitution, ma2024automated, liu2025research}, which could be mitigated by integrating hardware security modules into the corresponding controllers, requiring cryptographically signed execution tokens from certified sources. More broadly, automated research facilities may become targets for state or corporate espionage, and securing them may be vital for both biosecurity and the integrity of scientific markets.

Beyond physical dangers, automation raises bioethical concerns. An AI scientist optimising for efficiency might bypass necessary in vitro safety assays to proceed directly to in vivo animal testing. These moral boundaries are currently enforced by Institutional Review Boards (IRBs) and Institutional Animal Care and Use Committees, which may need to handle a substantially increased volume of AI-generated research proposals. Here, preliminary AI ethics assessments~\citep{fukataki2025developing} could provide automated compliance checks for low-risk, templated studies. Should the AI IRB detect a violation of clinical equipoise~\citep{emanuel2017makes}, failures to minimise harm, or absent consent provenance, the smart contract would void the protocol. Complex protocols that touch on sensitive areas would be escalated to human committees for key moral judgments~\citep{sridharan2024assessing}.

\subsection{Recursive Self-improvement}

AI science introduces the possibility of recursive self-improvement (RSI): AI agents autonomously conducting AI research to improve their own scaffolds or foundation models~\citep{nick2014superintelligence, yampolskiy2015artificial, yin2025godel, kang2026harnesscontinuallearningcontinual, chi2026ai4aibenchbenchmarkingllmagents}. The risk of RSI relates to instrumental convergence~\citep{benson2016formalizing} - regardless of an AI scientist's primary objective, certain instrumental sub-goals, such as self-preservation and resource acquisition are rational strategies for ensuring the primary objective is met. An unmonitored RSI loop introduces a significant risk of alignment drift~\citep{ngo2022alignment}. This drift could eventually erode bioethical constraints and other safety guardrails. Evals for "Autonomous Replication and Adaptation" (ARA)~\citep{kinniment2023evaluating} test for an agent's ability to autonomously acquire resources, spin up new server instances, and evade shutdown. Mitigating RSI risk requires embedding safety protocols directly into the evolutionary loop~\citep{sahoo2026sahoosafeguardedalignmenthighorder}, deploying AI research agents within strict sandboxes without external web access or the ability to independently acquire computational resources, and treating this research direction as a special case given the unique risks it presents.

\subsection{Open Science and Structured Access}

The dual-use and recursive risks discussed above may eventually conflict with the foundational scientific norms of openness and knowledge sharing~\citep{nosek2015promoting, nosek2012scientific}. Precedents for self-regulation exist. The 1975 Asilomar Conference on Recombinant DNA~\citep{berg1975summary} and the discussions surrounding the 2011 Dual-Use Research of Concern (DURC) controversies over gain-of-function H5N1 avian influenza experiments~\citep{fouchier2013transmission, duprex2015gain} demonstrate that dissemination may be restricted when faced with severe public health hazards. However, AI automation makes the DURC problem fundamentally asymmetric. Under the \emph{vulnerable world hypothesis}~\citep{bostrom2019vulnerable}, a technology enabling catastrophic destruction from widely available materials turns open science into a civilisational vulnerability, and AI scientists capable of rapidly generating new knowledge make the proliferation of dangerous information potentially irreversible~\citep{seger2023open}.

This dynamic may be seen through the lens of the offense-defense balance~\citep{garfinkel2021does}. Proponents of open-access rightly note that restricting access concentrates power and that broad access to capable AI agents may accelerate discovery of countermeasures~\citep{kapoor2024societal}. However, for certain risks (such as biological risks) the offense-defense balance may not favour openness, and acknowledging that unconditional scientific transparency may no longer be viable requires rethinking academic publishing, peer review, and institutional data-sharing norms.

One approach is to move towards a spectrum of \emph{structured access regimes}~\citep{shevlane2022structured}, facilitating widespread scientific collaboration while constraining the proliferation of dual-use capabilities. Since algorithms and data are easily copied once created, the most robust regulatory anchor for managing tiered access is compute governance~\citep{anderljung2023frontier, sastry2024computing} as the substantial resources required by autonomous AI scientists provide a natural enforcement point. Researchers and secondary AI agents could interact with AI scientists engaging in sensitive research exclusively via monitored cloud APIs, with access granted to trusted academic or government institutions and conditional on robust identity verification and detailed interaction audits. Unauthorised access or a dangerous intent would trigger instant revocation and tracing to human and institutional principals. For certain high-risk sensitive topics such as pandemic countermeasures, global collaboration may be necessary; for example, through international consortia. How such collaborations may be safely structured remains an open question, and the correct approach may vary on a case-by-case basis.

\section{International Collaboration and Value Distribution}
\label{sec:geopolitics}

Equitable distribution of value does not automatically follow from accelerated discovery. If AI science is left to market forces or state security mandates, value will concentrate within a few heavily capitalised corporations and technologically dominant states~\citep{bremmer2023ai}. Managing global collaboration, infrastructural ownership, and the distribution of scientific surplus requires new frameworks for international collaboration.

\subsection{International Collaboration}

Given the dual-use biorisks and AI RSI risks, the future of autonomous AI science may not be fully open. In the extreme, management of these risks may lead to national scientific silos. As states increasingly tie AI and advanced manufacturing to their national security and economic interest~\citep{lee2018ai}, they may restrict cross-border data flows and treat scientific discoveries as strategic assets rather than public goods~\citep{horowitz2018artificial}. The early stages of this fragmentation are already visible in the realm of \emph{genomic sovereignty}, where nations increasingly restrict the cross-border flow of population-level genomic data due to a mix of privacy concerns and strategic economic interests~\citep{majumder2016beyond, mckibbin2023genomic}. Initiatives such as the Canadian Sovereign AI Compute Strategy~\citep{ised_sovereign_ai_strategy_2026} and its associated AI Compute Access Fund~\citep{ised_ai_compute_fund_2026} represent direct state interventions designed to support national AI-driven scientific progress and retain the intellectual property and economic value of AI-driven discoveries within national jurisdictional boundaries.

Fragmentation of the scientific ecosystem would render it less efficient, potentially offsetting many benefits of automation. Siloing would impede building on advances achieved elsewhere and generate costly redundancy, both by duplicating successful lines of inquiry and by repeating expensive negative results already established in other jurisdictions. Large-scale cross-national challenges such as climate change and pandemic response further require close cooperation~\citep{gluckman2017science}. To mitigate these risks, international collaboration may need to shift from the direct exchange of assets to the exchange of vetted insights, supported by transnational, treaty-backed institutional frameworks~\citep{taeihagh2021governance, johnson2021survey, emery2025international}. Technological solutions can complement governance innovations. For instance, sovereign AI agents from two nations simultaneously developing a vaccine for a novel pathogen could collaborate without sharing sensitive genomic or proprietary data by utilising cDUAs, secure multi-party computation~\citep{zhao2019secure, knott2021crypten}, and federated learning methods~\citep{rieke2020future}. This could allow their respective agents to jointly train a shared predictive model via encrypted gradients while raw data never leaves its host jurisdiction. Establishing collaboration protocols that guarantee data sovereignty and algorithmic security may preserve the spirit of open science under conditions of geopolitical constraint.

\subsection{Compute and Capital}

International agreements must be paired with domestic economic policies to address the distribution of value derived from AI-driven science. In the traditional innovation economy, states act as primary risk-takers, funding early-stage foundational research, while private capital later translates these insights into commercial applications and captures the financial rewards~\citep{yencha2015entrepreneurial}. This model of \emph{public risk, private reward} is well established, particularly in biomedical research, where valuable therapeutics often trace their origins to public grants~\citep{galkina2018contribution}. The transition to autonomous AI science risks deepening these asymmetries, given the capital-intensive nature of the requisite compute infrastructure and that many frontier models are trained on scientific public commons - research papers, public biobanks, scientific software, and openly available experimental data. Without carefully crafted policies, AI scientists may enclose these publicly funded resources and convert them into proprietary intellectual property. 

Recognising the strategic role of compute, governments are beginning to invest in sovereign AI infrastructure and public compute reserves, such as the US National Artificial Intelligence Research Resource~\citep{parashar2023strengthening} and the European High Performance Computing Joint Undertaking~\citep{skordas2019toward}. However, if such reserves merely subsidise the overhead of private scientific operations without affecting IP ownership structures, the public may not see sufficient economic return. The economic mechanisms proposed in earlier sections - including tokenised hypothesis exchange and IP-NFTs - could encode public-interest clauses. For discoveries achieved via state-funded compute or data, smart contracts could automatically route a fraction of licensing royalties back to the sovereign compute infrastructure, creating a sustainable funding loop while mandating affordability for technologies of public interest.

Scaling public compute reserves to support large populations of autonomous AI scientists has consequences for energy consumption and grid capacity. Balancing the upside of automated discoveries against the carbon footprint of compute clusters will become an important policy consideration, and continued innovation in hardware efficiency will be necessary to mitigate these costs~\citep{wu2022sustainable, tamburrini2022ai, kirkpatrick2023carbon, luccioni2024environmental, luccioni2024power, luccioni2024light}.

\subsection{The Human Scientist}

Unlike previous waves of industrial automation that primarily displaced routine physical labour, the agentic scientific economy automates highly specialised cognitive work~\citep{autor2015there, webb2019impact}. If human cognitive labour is no longer the primary bottleneck to scientific progress, what is the role of the human scientist? Automation of routine scientific tasks does not render human cognition obsolete; instead, it drives human labour up the abstraction hierarchy~\citep{acemoglu2019automation}. Human scientists may therefore gradually transition from execution to orchestration and oversight. Current AI scientists, trained on historical data, remain better suited for incremental and combinatorial optimisation than for revolutionary, deeply intuitive leaps. Human experts are also necessary for making complex trade-offs and establishing ethical boundaries within their respective fields. The human scientist may therefore become the \emph{principal}, underwriting the actions of AI delegates, with human-driven research collaboration remaining the most credible deployment paradigm across the research lifecycle~\citep{kong2026aiautoresearchroadmap}. Yet for this model to work, human scientists need access to AI delegates, which requires computational resources. Stark disparities currently exist between researchers in different institutions. While the cost of running AI agents may decrease over time (increasing global accessibility of AI scientists), public compute grants or subsidies may be necessary to ensure that scientists across regions can access AI tools.

This model, however, presupposes a continued supply of expert human scientists, who develop their skills and intuitions through first-hand experience with challenging problems. Top experts are a minority of all trained researchers, and research \emph{taste} is hard to teach. If the AI-enhanced scientific economy assumes widespread availability of such talent to steer and oversee AI scientists, a natural question arises: where will this deep expertise come from in the future? Emerging empirical evidence suggests that habitual reliance on AI risks degrading human creativity~\citep{ashkinaze2025ai}. Universities and their research programmes compound this concern, as they play a dual role in society: they are centres of research excellence, but they are also educational institutions that produce both professional scientists and citizens with highly developed critical thinking skills. Over-reliance on AI ideation may erode both the depth of these skills and the intrinsic motivation that draws early-career scientists to research~\citep{wu2025human}. Should science automation disincentivise people from pursuing advanced degrees or substantially alter the nature of such training, the consequences may extend beyond the scientific workforce to broader social structures. 

Finally, there is a question of meaningful human participation in understanding. Science is ultimately a human endeavour whose purpose extends beyond producing goods or solving immediate social needs - it is how we uncover truths about the world and deepen our understanding. While traditional AI systems required bespoke solutions to extract AI-derived concepts~\citep{schut2023bridginghumanaiknowledgegap}, LLMs reason in natural language, enabling them to act as a semantic layer between specialised simulators, bespoke ML tools, and human collaborators. The challenge may therefore become one of volume rather than quality, provided that the AI discoveries map onto mechanistic models~\citep{posner2026observationinsightmechanisticworld} that lend themselves to accurate and insightful explanations. Should AI truly accelerate discovery, it may become infeasible for any human scientist to carefully review and deeply engage with a sufficiently large portion of the knowledge produced within their own field. This is likely to be among the most important challenges of advanced science automation.

\section{On the Limitations of AI-Driven Discovery}
\label{sec:epistemology}

To date, AI scientists have been limited in their ability to deliver profound conceptual leaps, raising concerns~\citep{zahavy2026position}. While AI scientists excel at recombinant innovation~\citep{fleming2001recombinant, weitzman1998recombinant}, by efficiently reusing known patterns and finding links between established concepts, such innovation may be more incremental. Though undoubtedly valuable for advancing what Kuhn formalised as \emph{normal science}~\citep{kuhn1970structure}, such reasoning is limited in its ability to generate foundational breakthroughs. Human science has seen a decline in disruptiveness~\citep{park2023papers}, and AI scientists should ideally help overcome this problem rather than contribute to or exacerbate it. It is likely that the capabilities of AI scientists will keep improving, with the increased availability of scientific learning environments and further advances in inference-time search. Work is also underway on improving the long-horizon planning of AI scientists~\citep{zhao2026scienceflowlonghorizonagentml} and on their applicability across modalities and disciplines~\citep{li2026omniscientist}. However, it may be prudent to complement AI scientist capability improvements with additional structural and sociotechnical interventions.

As science automation increases, AI scientists should carefully balance the amount of time spent on incremental work, and open-ended creative discovery - driven by intrinsic motivation~\citep{chentanez2004intrinsically, barto2012intrinsic, kulkarni2016hierarchical, achiam2017surprise, aubret2019survey} and continuous re-examination of prevailing theories and practices. Early examples of how intrinsic motivation may be incorporated in AI scientists can be seen in recent applications in formal mathematical reasoning~\citep{NEURIPS2024_4b8001fc}.

The utilisation of LLMs in AI scientists introduces the risk of \emph{cognitive monoculture}~\citep{fazelpour2026navigatingepistemicmonoculturesaidriven}, as well as \emph{citation monoculture}~\citep{alemohammad2026aiwritesgetscited}. This is of particular concern given that preliminary findings point towards a reduction in idea diversity with the use of generative AI in scientific workflows~\citep{hao2026artificial, meincke2025chatgpt, traberg2026ai}. Cognitive monoculture may also emerge in human research communities, especially as conservative grant funding tends to reward low-risk proposals that do not challenge the status quo~\citep{schweiger2024costs}. Yet, a degree of pluralism tends to be highly beneficial, and sometimes strictly required, especially when research intersects with ethics~\citep{chandak2026does}. No consensus exists on the extent of cognitive monoculture risks in science, though it is widely recognised that it may lead to overlooking promising research directions. This risk is lower in well-established sub-fields, presuming that the assumptions and edge cases are properly recognised~\citep{hedden2026algorithmicmonoculturecritics}. In the history of science, many now famous papers have initially been rejected by peer review, as they were challenging established research dogma. Retaining the ability to effectively challenge dogma is therefore critical.

More research is needed to understand if AI scientists systematically prioritise certain types of ideas, as a prerequisite for designing effective interventions. Given that AI scientists are often trained on domains where both a wide availability of tools and a strong, efficiently computable reward signal exists, this could lead to a form of the McNamara fallacy~\citep{jannach2020escaping, bisht2026agenticaiscientistsbuilt} where easily quantifiable directions are strongly prioritised irrespective of their intrinsic value. Human oversight mitigates these risks, though this hinges on active participation and cognitive friction. As Messeri and Crockett \citep{messeri2024artificial} highlight, the uncritical reliance on highly capable AI tools fosters an \emph{illusion of understanding}. Individual researchers may also lack the field-wide overview needed to notice the absence of less common viewpoints in the output. It may therefore be advisable to complement the individual human steering with structural interventions for inducing strategic heterogeneity~\citep{repec:arx:papers:2604.09502}. Diversity objectives embedded within the scientific idea markets may be designed to reward and allocate resources towards non-consensus research directions. The desire to maintain a diversity of scientific ideas would need to be balanced against the practicalities of identifying and discontinuing non-promising directions, as well as the overall cost and efficiency.

Finally, it is important to consider the risk of introducing epistemic opacity~\citep{creel2020transparency, humphreys2004extending} through increased use of AI scientists. Epistemically opaque models can simultaneously be highly productive within the \textit{context of discovery}, while also introducing a severe liability in the \textit{context of justification} through their opacity~\citep{Duede_2023}. These concerns have been partially reflected in the discussions following the AI-derived solution to the Millennium Prize Navier-Stokes problem in mathematics, resulting in a declaration signed by some of the world's leading mathematicians, namely \emph{A Severe Misalignment of AI in Mathematics}~\citep{avila_2026_22737751} - serving as a warning that mass automation of proofs in mathematics may be at odds with the core objective of the field, which is to advance human understanding and build new conceptual frameworks.

This risks creating a critical bottleneck in future scientific workflows. Human institutions are responsible for allocating physical resources, setting constraints, and assuming legal liability for scientific experiments. Human researchers must therefore be able to act according to this responsibility, and comprehend and audit the rationale underpinning AI proposals and action plans. Science is inherently a collaborative, social endeavour~\citep{channing2026ai}, where a black-box oracle would alter the nature of scientific understanding~\citep{sullivan2022understanding}, even if highly accurate. It would be preferable if AI scientists could instead be utilised as interpretable reasoning engines~\citep{truhn2023large}, surfacing the causal mechanisms behind their proposals~\citep{kiciman2023causal}. Interpretability helps co-creation, by enabling researchers to audit AI reasoning, set constraints, and align automated discoveries with societal objectives.

\section{Conclusion}
\label{sec:conclusion}

The role of AI within scientific research is rapidly evolving. Early AI systems were useful predominantly as highly specialised tools, whereas new general-purpose systems and AI agents are enabling the development of AI scientists capable of autonomously or semi-autonomously pursuing different types of research tasks. This transition raises the question of whether existing scientific infrastructure is suited for a future in which AI scientists can autonomously pursue scientific discovery. While this is not yet the case, we should be actively considering the necessary changes, to prepare societies to reap the benefits of this hypothesised acceleration.

We argue that the development of AI scientists is likely to be bottlenecked primarily by physical resources and empirical validation, rather than the ability to produce plausible or promising research ideas. It is the economic, legal, and physical friction that may stand in the way of realising the downstream benefits. With that in mind, to avoid future market failures, the scientific community must proactively develop a native Automated Scientific Economy. This economy would make trade-offs between scientific pursuits explicit, include mechanisms to represent the public interest, and would help avoid blind spots and neglected topics. It would also make it possible to financially decouple the computational labour of ideation from the capital-intensive labour of physical execution, and would enable agents and institutions to negotiate priority access to limited physical resources depending on the estimated value of different research proposals. Markets would also require solutions for appropriate pricing and sharing of negative results, secure computational approaches for handling sensitive data within trusted execution environments, IP frameworks, and information security risks. Deep macroeconomic interventions may be required to establish the appropriate public compute reserves and subsidies, coupled with bespoke policies, regulations, and legal frameworks. As more tasks are delegated to AI scientists, human experts may shift towards orchestrating AI teams, interpreting results, and maintaining safety and ethics standards.

\section*{Disclaimer}

The opinions presented in this paper represent the personal views of the authors and do not necessarily reflect the official policies or positions of their organisations.

\bibliography{main}

\end{document}